\documentclass[preprint]{JHEP3} % 10pt is ignored!

\JHEPspecialurl{http://jhep.sissa.it/JOURNAL/JHEP3.tar.gz}

\usepackage{epsfig,multicol}

\def\beq{\begin{equation}}

\def\eeq{\end{equation}}

\def\CA{{\cal A}}
\def\CB{{\cal B}}

\newcommand{\bea}{\begin{eqnarray}}

\newcommand{\eea}{\end{eqnarray}}

\parskip=\itemsep               %?

\newcommand{\beqar}[1]{\begin{eqnarray}\label{#1}}

\newcommand{\eeqar}{\end{eqnarray}}

\newcommand{\om}{\omega}

\def\thefootnote{\fnsymbol{footnote}} 
\title{
{\Large  \bf Improved Hydrodynamics from the AdS/CFT}}
\author{M. Lublinsky and E. Shuryak   \\
Physics Department, SUNY Stony Brook,  NY 11794-3800, USA}

\abstract{
We generalize  (linearized) relativistic hydrodynamics  by including all order gradient expansion
of the energy momentum tensor,
 parametrized by four momenta-dependend transport coefficients,
one of which is the usual shear viscosity. We then apply the AdS/CFT duality for ${\cal N}=4$ SUSY 
in order to compute the retarded correlators of the energy-momentum tensor. From  these correlators
we determine a large set of transport coefficients of third- and fourth-order hydrodynamics.
  We find that higher order terms have a tendency to reduce  the effect of viscosity.
\\
}

\begin{document}

%*********************************************************************************  

%*********************************************************************************  

\def\thefootnote{\arabic{footnote}} 

\section{Introduction}

Relativistic Heavy Ion Collider (RHIC) produces 
hadronic matter  with temperatures ranging between
the initial  $T_i\sim 2T_c$ and the final (or freezeout) value $T_f\sim T_c/2$ \cite{Rom,T2},
where $T_c \approx 170 MeV$ is the QCD critical temperature.
 It has been shown \cite{Teaney:2001av,Hirano:2004ta,Nonaka:2006yn}
that the flows (radial, elliptic) associated  with the plasma expansion are well and consistently described by 
near-ideal relativistic hydrodynamics, with freezeout implemented via hadronic cascades. 
Since elliptic flow is dominated by the early times of the fireball expansion, when $T>T_c$ and matter is
in the so called Quark-Gluon Plasma (QGP) phase \cite{QGP}, this lead to the conclusion
\cite{Gyulassy:2004zy,Shuryak:2004cy} that QGP  is a ``perfect liquid'', 
presumably    because it is actually  in a new -- strongly coupled -- regime of QCD.
Those views were discussed in detail and eventually accepted in the 2004 ``white papers" of all four  experimental collaborations \cite{wp}.
The full understanding of QCD dynamics
 at strong coupling, even in the deconfined phase, remains a challenge and one usually appeals to either lattice simulations
or phenomenological models. While the lattice  is considered 
 a reliable source for QCD thermodynamics, it usually fails to provide accurate data
on transport coefficients. Thus, in order to understand transport properties of QCD, at the moment we have 
to appeal to various microscopic models (for recent reviews on the strongly coupled
QGP see	e.g.\cite{Edrev, ST}). 

Starting from 2004  RHIC experiments have discovered and studied
phenomena known as the ``cone'' and the ``ridge'', associated with the propagation
of the energy deposited by quenched hard jets (for the  description of the  phenomena 
and recent data see Refs. \cite{:2008nda,Ajitanand:2006is,2009id} 
and references therein).   
The former  was associated with conical hydrodynamic flows induced
by fast particles propagating through the medium \cite{conflow}. This has in turn initiated studies, within 
 AdS/CFT, of such  processes  induced by a steadily moving heavy quark, see \cite{CY,Yarom,Gubser,Gul}. 
Although these studies did not appeal to any 
hydro description, their results were found to be in very good agreement with (even unimproved) hydrodynamics.

 Another structure, known
as ``soft ridge" has been observed in two-particle correlations: its origin
is attributed to initial state fluctuations in the colliding nuclei.
For experimental data and phenomenological discussion see the talks at the BNL dedicated workshop \cite{cathie_workshop}.
 Although in this paper we will not
 discuss the phenomenology of those objects, we  nevertheless stress that they provide
the strongest motivation
for a detailed study of small perturbations on top of (hydrodynamically expanding) matter.

 One obvious  step in understanding these perturbations is to study
them using  {\em linearized} hydrodynamics, assuming their amplitude to be small.
 On the other hand, these objects start their evolution at 
much smaller scales
compared to nuclear (or fireball) radii. For example initial state fluctuations
are believed to be given by the  ``saturation scale" $1/Q_s$, which is only $~0.2$ fm, about 30 times smaller than the fireball as a whole.
Therefore, the evolution of small perturbations includes much larger spatial gradients, and
in order to treat them better
one would naturally try to improve the accuracy
of  hydrodynamics, including {\em higher order derivative terms}.
Since the latter  appear with many new transport coefficients,
the usual phenomenological approach which derives viscosity
from the data  would hardly be possible. Instead, some 
self-consistent approach
is needed to calculate as many of them as necessary.

Such model of choice for the present study is   
${\cal N}\,=\,4$ SUSY at large $N_c$. Via the celebrated AdS/CFT correspondence \cite{Malda}
this gauge theory at strong coupling admits a dual description in terms 
of  weakly coupled gravity in AdS$_5 \ \otimes$ S$_5$ space.
  The finite temperature version of this field theory
is dual to the AdS-Schwarzschild black hole (brane). The 
laws for Schwarzschild black hole thermodynamics
imply that the entropy density is proportional to the area of the horizon 
\cite{Klebanov_etal}
\footnote{For theories involving higher order curvature corrections to the Einsteinian gravity,
the relation between area of the horizon and  entropy is invalid \cite{Myers}.}.
The equilibrium pressure is $P\,=\,\pi^2\,N_c^2\,T^4/8$, while the energy density is $\epsilon\,=\,3\,P$ 
due to the conformal symmetry of the microscopic theory.
  
  Refs.  \cite{KSS} pioneered the study of transport coefficients via dual description. For a
static plasma and in the limit of large 't Hooft coupling $\lambda\gg1$, %the authors of \cite{KSS}
 %found that
  the ratio of shear viscosity to entropy is independent of the coupling and is in fact remarkably small 
\beq\label{bound}
{\eta_0\over s}\,=\,{1\over 4\,\pi}\,.
\eeq
Furthermore, Refs. \cite{KSS} conjectured that this value for the ratio
is a universal lower bound, valid for all physical systems in nature. 
While  AdS/CFT leads Eq. (\ref{bound}),
 it does not provide any explanation of this result from the gauge theory side.
So far, no microscopic mechanism for low viscosity  has been established for QCD, though
a promising proposal  can be found in \cite{clad}.

It 
is rather difficult to extract viscosity from experimental data precisely, because  it is small and its effects are
$O(10\%)$ or so, comparable to  other uncertainties.
The phenomenological studies of  RHIC data (such as in Ref \cite{RR})
 typically focus on the elliptic flow dependence on centrality or transverse momentum, $v_2(b,p_t)$. 
The optimal $\eta_0/s$ for these fits occurs at a value of the order of the suggested minimum, although deviations
from it by a factor two or so are still possible.
Another argument to support very low viscosity comes from discussions of the overall entropy production, such as in
Ref. \cite{Dum}. Those works suggested that there is 
a tension between the total entropy (measured by the observed multiplicity of produced hadrons) and the 
very short thermalization time (initial time for hydro evolution),
unless the viscosity over entropy ratio is pushed down, maybe even below the bound. 
The third (more indirect) argument for low viscosity is the survival till freezeout 
of the ``cones" and ``ridges'', suggesting smallness of  dissipative effects.
  Therefore all these approaches indicate a  very small viscosity value.

The hydrodynamic representation of the energy-momentum tensor is
\beq
\langle T^{\mu\nu}\rangle\,=\,(\epsilon\,+\,P)\,u^\mu\,u^\nu\,+\,P\,g^{\mu\nu}\,+\,\Pi^{\mu\nu}
\eeq
where the average is taken over the thermal bath. While at some microscopic scale $l$
the system is assumed to be locally at thermal equilibrium, at some macroscopic scale $L\gg l$ the
 local fluid velocity field $u$ is a function of space-time coordinates. 
The ``tensor of dissipations" $\Pi^{\mu\nu}$, added to the ideal-fluid part, represents all the deviations from the
equilibrium state induced by such a flow field.  In the long wavelength  limit   $L\gg l$,
these fluctuations can be expanded in terms of gradients of the velocity field, or in powers of $l/L$.
 The first order Navier-Stocks (NS) 
hydrodynamics retains only the first gradient \footnote{Throughout this paper we will be considering conformal theory only,
for which there is only shear viscosity since the bulk viscosity is zero.}
\beq
\Pi^{\mu\nu}\,\sim\,\eta_0\,\nabla^\mu u^\nu\,.
\eeq
In this work we will discuss higher order gradients, which will provide certain corrections to the first order
viscosity term when gradients grow. These corrections are relevant for smaller size objects in the plasma or to earlier
time of hydro evolution.

The high order gradient expansion generically includes two types of terms: (i) non-linear terms in the velocity field 
(like $(\nabla u)^2$) and (ii) linear terms
with multiple gradient operators acting on a single velocity field (like $\nabla\nabla u$). 
These two types of terms are controlled by two different
parameters. The non-linearities are important when the field amplitude is large. However, even for small amplitude waves, 
one can get large contributions from  the linear terms when the 
momenta associated with the wave are large.

 Recently, second order hydrodynamics (next-to-NS) attracted significant attention \cite{Rom,T2}. The main reason is
 that  NS hydrodynamics is known to have causality problems. 
%The acausal effects
%lead to such unphysical phenomena as spontaneous self-heating and entropy decreases. 
The accausal effects  create
numerical instabilities
when solving  hydrodynamic equations. The problem originates from
the fact that
 NS equations imply  instantaneous response to any perturbation
introduced in the system.
%In this paper we will be interested in these higher order gradients, but will restrict ourselves to the linear terms only.
%%re usually not very well equilibrated and the finite momenta effects might be important.
%We will elaborate on this point at the end of this section.
In order to circumvent this problem, one may introduce a relaxation time.
It explicitly appears as a new transport coefficient when the gradient expansion
is extended to second order:
\beq\label{sec}
\Pi^{\mu\nu}\,\sim\,\eta_0\,[1\,-\,\tau\,(u\,\nabla)\,]\,\nabla^\mu u^\nu\,
\rightarrow\,\eta_0\,[1\,+\,i\tau\,\om]\,(k^\mu\,u^\nu)\,.
\eeq
In fact, in order to restore causality it is not sufficient to include second gradients only:
all order gradients need to be resummed. A very popular resummation scheme
is due to Israel and Stewart (IS) \cite{IS}. It essentially generalizes  viscosity 
to an $\om$-dependent  but $k$-independent (complex) function
\beq\label{IS}
\eta^{IS}(\om)\,=\,{\eta_0\over 1\,-\,i\,\om\,\tau}\,.
\eeq
Eq. (\ref{IS}) can be viewed as a Pade-like resummation of (\ref{sec}). The relaxation time $\tau$
provides  a scale for  exponential relaxation. The position of the pole below the real axis
and the ``good'' falling off asymptotic behavior make the model causal. In other words, as a function
of complex $\om$, the viscosity 
is analytic in the upper half plane.
In coordinate space this simple pole corresponds to a  memory function with exponential falloff:
\beq
\Pi^{\mu\nu}(x,t)\,\sim\,{\eta_0\over \tau}\,\int_0^t\, dt^\prime \,e^{-(t-t^\prime)/\tau}\ 
\nabla^\mu\,u^\nu(x,t^\prime)\,.
\eeq
 In this paper
we will be studying all order gradient expansion in the linear approximation. Instead of introducing
new transport coefficients at each new order, we will be thinking of viscosity  and other transport 
coefficients as frequency and momentum dependent functions. We will be working 
in the framework of ${\cal N}=4$ SUSY. For the rest of this paper
we set all dimensionfull units to be related with the temperature,  $2\,\pi\,T\,=\,1$. %All dimensionfull quantities are to be  measured in these units.
Among our results, we will show that the IS resummation, although well-known and used, is
 still simplistic model for high order gradient terms, which is even qualitatively inconsistent with AdS/CFT results.
Not only it misses important  non-linear terms already at second order  \cite{BSSSR,Minwalla},
but (as we will show below) it is also incorrect in the linear approximation starting from the third order.

%Poles of the viscosity go into relaxation scales, in time and space. 
More generally, we will find that higher order terms do have a tendency to cancel (or reduce) the effect
of  NS viscosity. In particular, in our earlier paper
 \cite{LS}, we argued that the extremely low viscosity suggested by Refs. \cite{RR,Dum} may essentially 
be some ``effective viscosity'', which includes  these high order gradient terms. 
The real systems probed in RHIC collisions have finite gradients and the inclusion
of their effects may demand  going 
beyond  NS approximation. 
In \cite{LS} we attempted to extract a momentum-dependent viscosity from the imaginary part of the sound dispersion curve.
Our main observation was that the effective viscosity as probed at finite momenta turns out to be smaller compared to the value at the origin. 
Motivated by \cite{Dum}, we discussed in \cite{LS}
the implications of a momentum-dependent viscosity on the entropy production for  Bjorken expansion
\cite{Bj}.
We discovered that the inclusion of  momentum-dependence made it possible  to push the hydrodynamic
 description a bit further into  earlier
times of the collisions, with 
 the entropy production due to viscous hydro  stabilized at around 20\%
of the total entropy produced in the collision. 
The conclusion is that the account for a momentum-dependent viscosity
reduces the sensitivity to thermalization time.
Now, with the result  reported below,  our previous approach \cite{LS}  based on  the sound dispersion
curve looks rather  naive (for a much more elaborated study of  hydrodynamic theory as an
effective theory for the lowest modes see Ref. \cite{Amado}). 
In general, we will see that the sound dispersion curve does not contain 
enough information  to define the ``generalized viscosity" function.  
Nevertheless, we qualitatively captured at least the right trend:  full second
order hydro with all non-linear terms included has the same trend towards reducing  the 
entropy production \cite{BSSSR,Heller}.

Our goal in this paper is to put the idea of a momentum-dependent viscosity
on a more solid ground compared to our naive treatment in \cite{LS}. 
In the present
analysis we will be focusing  on the retarded correlators of the stress tensor. 
The correlators contain information not only about the positions of the poles but also about their
residues. The complete information on the correlators is equivalent to the 
knowledge of the energy momentum tensor in the linearized approximation.

In a conformal theory in four dimensions,  there are only three independent correlators
of the energy-momentum tensor. 
These are correlators in the sound ($G^S$), shear ($G^D$), and scalar ($G^T$) channels.
AdS/CFT correspondence provides a tool to compute these correlators by solving certain
linearized gravity equations in the background of the AdS-Schwarzschild black hole \cite{PSS,Kovtun,SS}. 
These equations essentially describe graviton's propagation from the AdS boundary,
where the field theory is defined, to the horizon of the black hole. Absorptive boundary conditions are 
imposed there. Dissipation takes place at the horizon while there is no dissipation in the bulk of  AdS.  
However, the bulk curvature acts as a non-linear medium, which provides a source for complicated dispersion.
It is this dispersion which, by means of the duality,  is mapped  into momenta-dependent transport coefficients.

Our strategy is to first write a most generic hydro-like representation
of the energy momentum tensor $T^{\mu\nu}$, in terms of the
fluid velocity field $u$. We find that, generically, there are four structures (or operators involving derivatives
of $u$ or the metric $g$) which can occur in $T^{\mu\nu}$ and are consistent with all  symmetries. Each structure 
enters with a coefficient which is momentum-dependent. These are the generalized transport coefficient 
we are looking for. One of them is associated with the shear viscosity, while the remaining three
encode responses of the system to external (4d) gravity perturbations. We call them gravitational susceptibilities
of the fluid (GSF). The operators which are multiplied by the GSFs involve the Weyl tensor of the metric
and vanish in the flat Minkowski space.

We the proceed by using this hydro-like representation of  $T^{\mu\nu}$ in order to compute its correlators
in the three channels introduced above. We then attempt to determine the momentum-dependent 
transport coefficients from the matching to the functions
$G^S$, $G^D$, and $G^T$ computed directly from the bulk gravity side.

Our program runs into a problem, which we were not able to resolve completely:
there are in fact  four independent transport functions to be extracted from three equations.
Despite the fact that we could not determine the entire functions, we were able to
get them to quite high order in the perturbative expansion at small momenta.    
In particular, we found the shear viscosity function to fifth order  in the gradient expansion.
This involves several new transport coefficients, most of which are obtained numerically.

The conceptual problem mentioned above, prevented us from computing  shear viscosity  in 
the whole kinematic region of arbitrary frequency and momentum. Instead, we build a model similar to IS
which utilizes the information about the new transport coefficients and preserves the 
causality condition. We propose 
this model for phenomenological studies of hydrodynamics at RHIC, but 
any  application of this model is left beyond the scope of this paper.

The paper is organized in the following way.
In Section 2 we present the general setup for computing the retarded correlators from the bulk gravity and from the 
generalized hydro on the boundary. Section 3 presents some results. A phenomenological model for generalized viscosity
is proposed in Section 4. Our conclusions are summarized in Section 5. 
Two Appendices supplementing Section 2
provide details of some analytical computations.

\section{Generalities}

The retarded correlators of two energy-momentum tensors are defined as follows
\begin{equation}
  G^{\mu\nu \alpha\beta} (k,\om)\,
  = \,-\,i\,\int_0^\infty dt\,\int d^3x \,e^{-i\,\om\,t\,+\,i k x}\,
  \langle[T^{\mu\nu}(x,t),\, T^{\alpha\beta }(0)] \rangle \,
\label{retarded}
\end{equation}
Here the average is over the equilibrated thermal bath. 
For conformally invariant plasma with traceless $T^{\mu\nu}$,
 there are only three independent correlators
$G^T\equiv G^{xyxy}$ (tensor), $G^D\equiv G^{txtx}$ (shear),
and $G^S\equiv G^{tztz}$ (sound) with the vector $k$ pointing in the $z$-direction.
All other correlators are related to these three
either by  rotational symmetry or by the equations of motion.
\subsection{Life in the bulk: Retarded correlators from 
gravity}

In this subsection we closely follow the setup and results of Ref. \cite{KS}.
From the bulk gravity side,  in order to compute the retarded correlators at non-zero temperature
one has to solve certain wave equations (one for each
symmetry channel).  These equations describe propagation of the corresponding
metric perturbations  (gravitons) in the
AdS-Schwartschild BH background of the dual description.
The differential equations are of the form
\begin{equation}
  \frac{d^2}{dr^2} Z_a(r) +
  p_a(r) \frac{d}{dr}Z_a(r) +
  q_a(r) Z_a(r) = 0 \,,
\label{eq:master-equation}
\end{equation}
where the coefficients $p_a(r)$, $q_a(r)$
depend on the frequency $\om$
and momentum $k$, and $a=T,D,S$ labels the three symmetry channels.
The coefficient functions are given by the following expressions.

\noindent $\bullet$ The scalar channel
\begin{equation}
  p_T(r) = -\frac{1+r^2}{rf}\ , \ \ \ \ \ \ \ \ \ \ \ \ \ \ \ \ \
  q_T(r) = \frac{\om^2 - k^2 f}{r f^2}\ ,
\end{equation}
where $f=1-r^2$. The function $f$ is inherited from the AdS-BH metric. 

\noindent $\bullet$ The shear channel
\begin{equation}
  p_D(r){=}\frac{(\om^2-k^2 f)f + 2 r^2\om^2}{rf(k^2 f-\om^2)}\,, \ \ \ \ \ \ \ \ \ \ \ \ \ \ \ \  
  q_D(r){=}\frac{\om^2 - k^2 f}{rf^2}\, .
\end{equation}

\noindent $\bullet$ The sound channel
\begin{eqnarray}
  p_S(r) &=& -\frac{3\om^2 (1+r^2) + k^2 ( 2r^2 - 3 r^4 -3)}
           {r f (3 \om^2 +k^2 (r^2-3))}\ , \nonumber \\ && \nonumber \\
  q_S(r) &=&  \frac{3 \om^4 +k^4 (3{-}4r^2{+}r^4) +
            k^2 (4r^2\om^2{-}6\om^2{-}4r^3 f)}
            {r f^2 ( 3 \om^2 + k^2 (r^2 -3))}\, . 
\end{eqnarray}
The fifth dimension coordinate $r$ ranges from $0$ to $1$,
where $r=0$ corresponds to the boundary of the 
asymptotically AdS space, and
$r=1$ corresponds to the event horizon of the background metric.

The information about the retarded correlation functions
is encoded in the solutions to Eq.~(\ref{eq:master-equation}),
which satisfy the incoming wave condition
at the horizon $Z_a(r\rightarrow 1)\sim exp[-i\om/2]$. At $r=0$
the solution can be written  as a
linear combination of two independent local solutions,
\begin{equation}
  Z_a(r) \,=\, \CA_{a}\, Z_a^{I}(r)\, +\, \CB_{a} \,Z_a^{II}(r) \,,
\label{eq:Z-two-solutions}
\end{equation}
Here  $Z_a^{I}$ is irregular in the origin while  $Z_a^{II}$ is a regular solution.

The prescription to compute the correlators $G$ follows from the Minkowski formulation of the AdS/CFT
correspondence and amounts to computing the ratio between the two coefficients in the expansion 
(\ref{eq:Z-two-solutions})
\begin{equation}
   \tilde G^a(\omega,k) \,= \,-\,8\,P\,
    \frac{\CB_{a}(\omega,k)}{\CA_{a}(\omega,k)}\,.
\end{equation}
For the three symmetry channels the correlators $G$ are related to 
$\tilde G$,
\beq
G^{xyxy}\,=\,{1\over 2}\,\tilde G^T\,; \ \ \ \ \ \ \ \ \ \ \ 
G^{txtx}\,=\,{1\over 2}\,{k^2\over \om^2\,-\,k^2}\,\tilde G^D\,
\eeq
For the sound channel the relation is a bit more involved and includes a contribution from contact terms
\cite{KSqnm}
\beq
G^{tttt}\,=\,{1\over 2}\,\left[{4\over 3}\,{k^4\over (\om^2\,-\,k^2)^2}\,\tilde G^S \,+\,{1\over 12}\,{29\,k^4\,-\,30\,k^2\,\om^2\,+\,
9\,\om^4\over (k^2\,-\,\om^2)^2} \right]
\eeq

Eq. (\ref{eq:master-equation}) has real coefficients which are even 
functions of frequency. In other words this equation propagates waves
without any dissipation. The dissipation (time irreversal) effects are 
introduced by the boundary conditions at the horizon. 
However, the AdS-BH metric acts as a non-linear medium for the propagating 
graviton. The non-linear dependences on frequency and momenta which appear
 in (\ref{eq:master-equation}) are to be mapped onto highly non-trivial momenta dependence
of the transport coefficient functions.

Let us make a technical remark on numerical solution. The equations for
the shear and sound channels have singular points inside the bulk $r=[0,1]$.
For the shear channel it appears for $\om < k$ at $r_0=\sqrt{1-\om^2/k^2}$ 
and for the sound it is at $r_0=\sqrt{3\,(1-\om^2/k^2)}$ (condition that $r_0$ is inside the bulk).
It would be interesting to understand if these points have any special physical role.
To ensure that there is no instability caused by these singularities, we split our numerical solution
into two intervals $[0,r_0]$ and $[r_0,1]$ and matched the solutions at the singular points
using  analytic solutions in the vicinity of $r_0$.

The correlators computed from the gravity side
agree with the field theory correlators up to a constant \cite{PSS}.
In particular, in the sound channel the relation between the correlators $G^{tttt}$ and $G^{tztz}$ is
\beq\label{shift}
\om^2\, (G^{tttt}\,+\,\epsilon)\,=\,k^2\,(G^{tztz}\,+\,P)
\eeq

The analytical expansion for the correlators at small momenta can be found in Appendix A.
For the shear and sound channels we show some numerical results alongside the 
corresponding curves for the NS and IS hydrodynamics on Figs. \ref{fig1} and \ref{fig2}.
\FIGURE{\epsfig{file=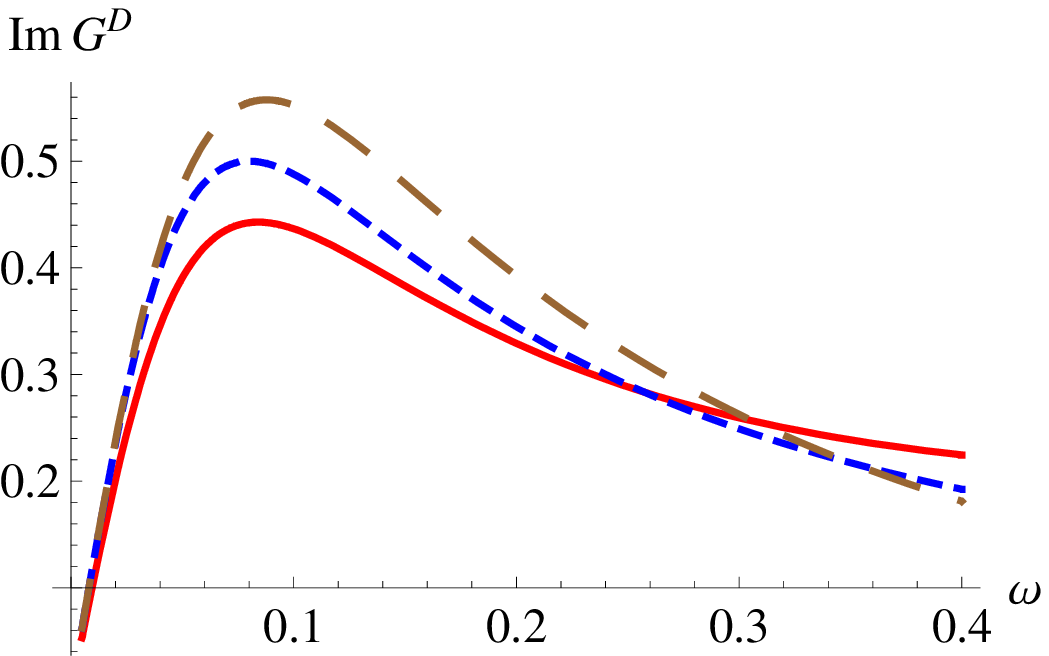,width=80mm} ~
\epsfig{file=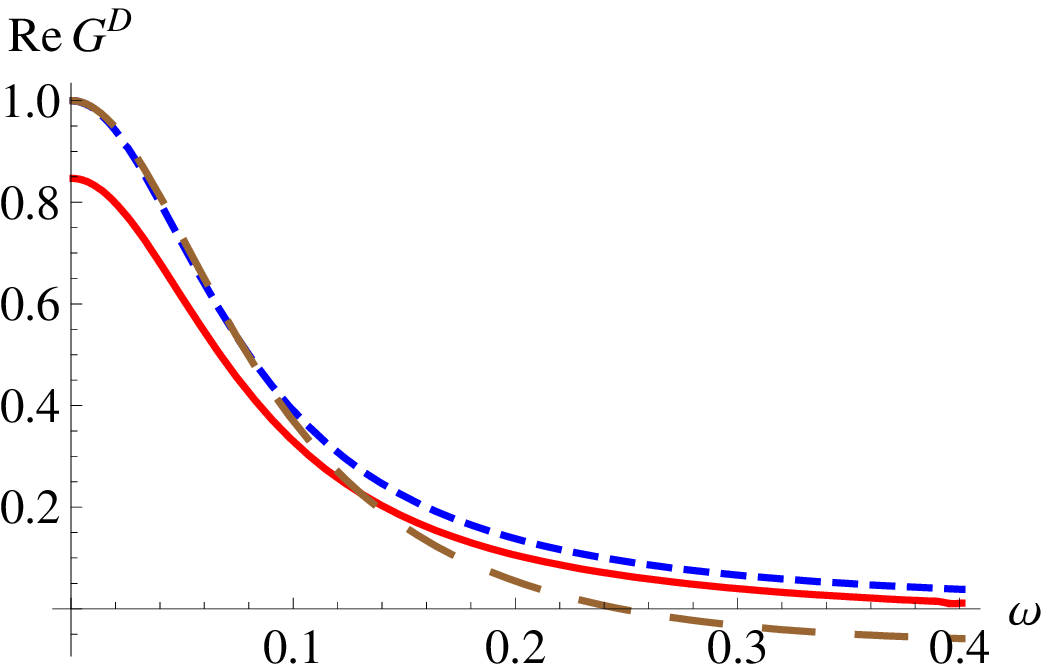,width=80mm} \\
\epsfig{file=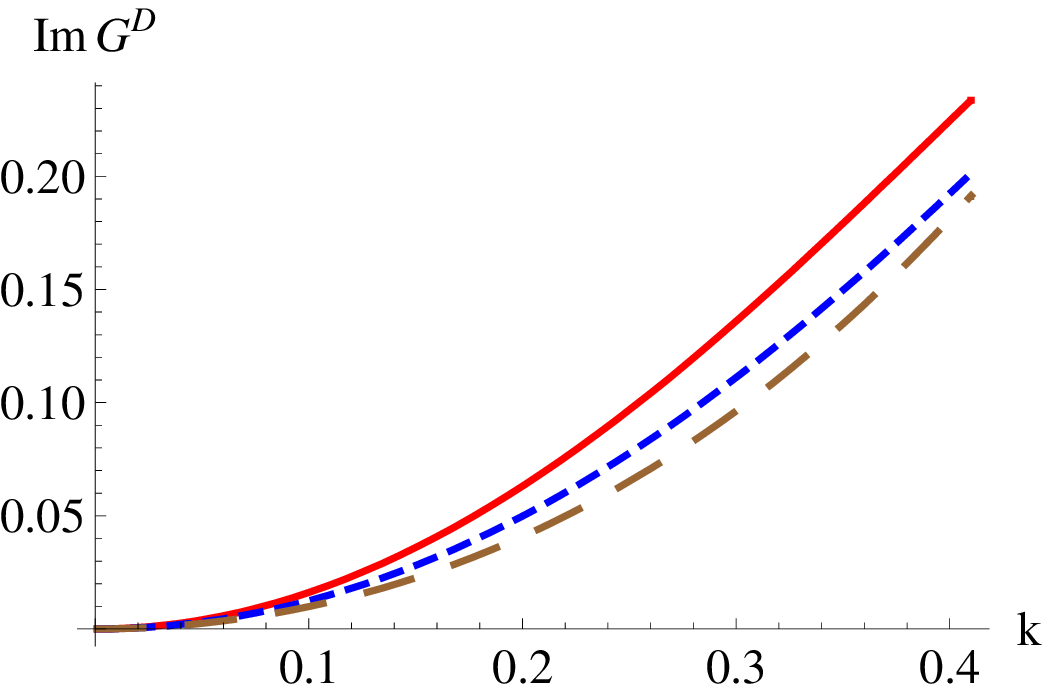,width=80mm} ~
\epsfig{file=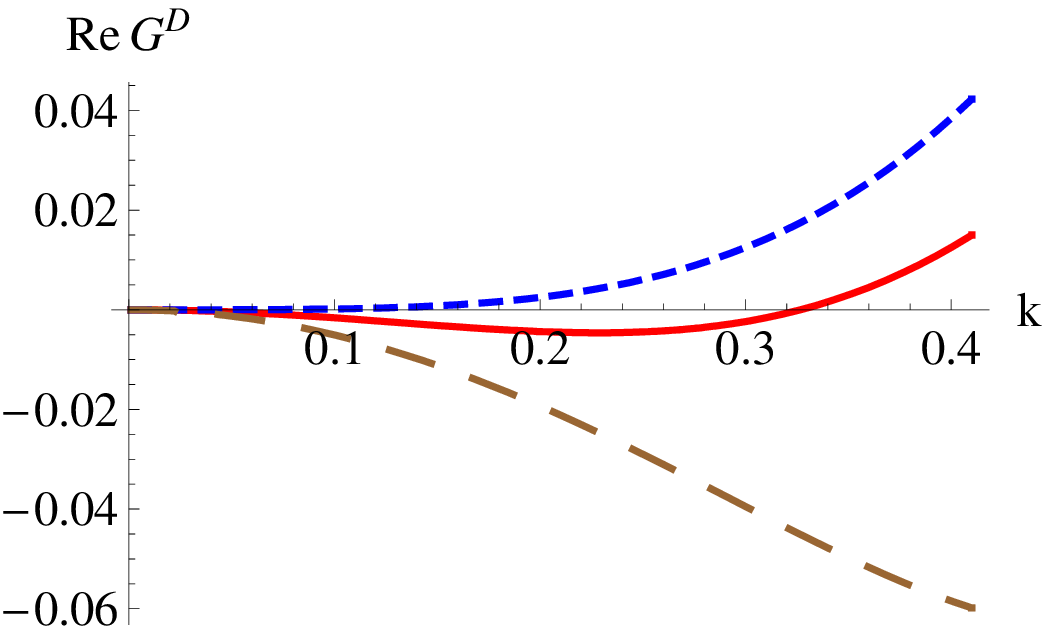,width=80mm}
\caption{\it Shear channel: top $k=0.4$; bottom $\omega=0.4$.
Solid  line corresponds to the AdS/CFT correlator. Short dashes display the NS hydrodynamics while
long dashes show the IS hydrodynamics.}
\label{fig1}
}
\FIGURE{\epsfig{file=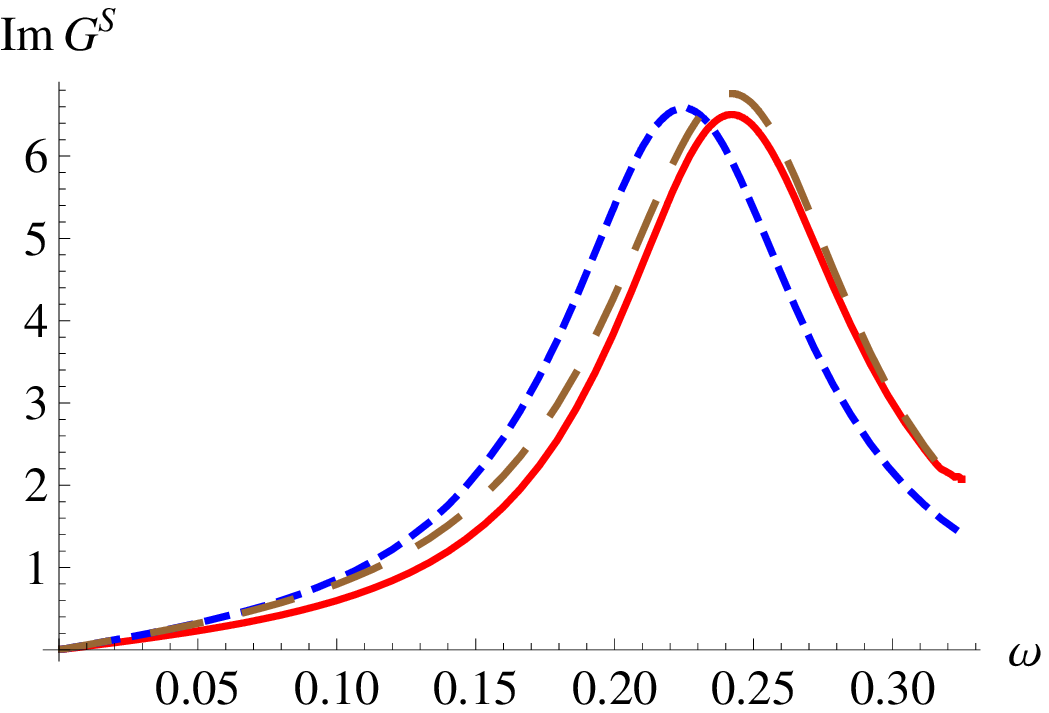,width=80mm} ~
\epsfig{file=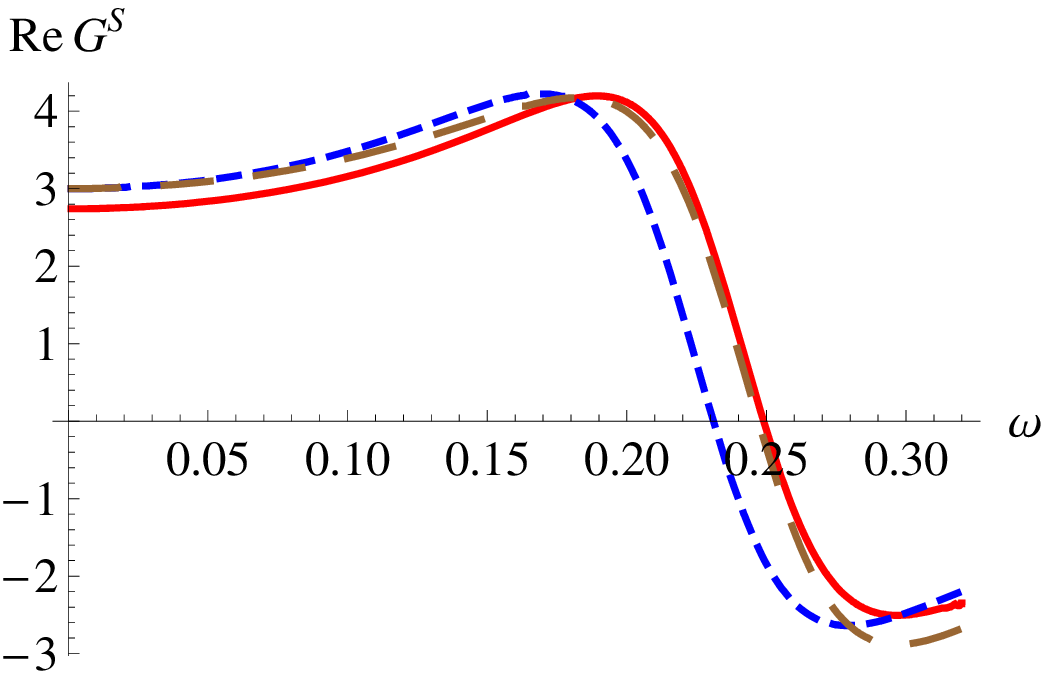,width=80mm}
\epsfig{file=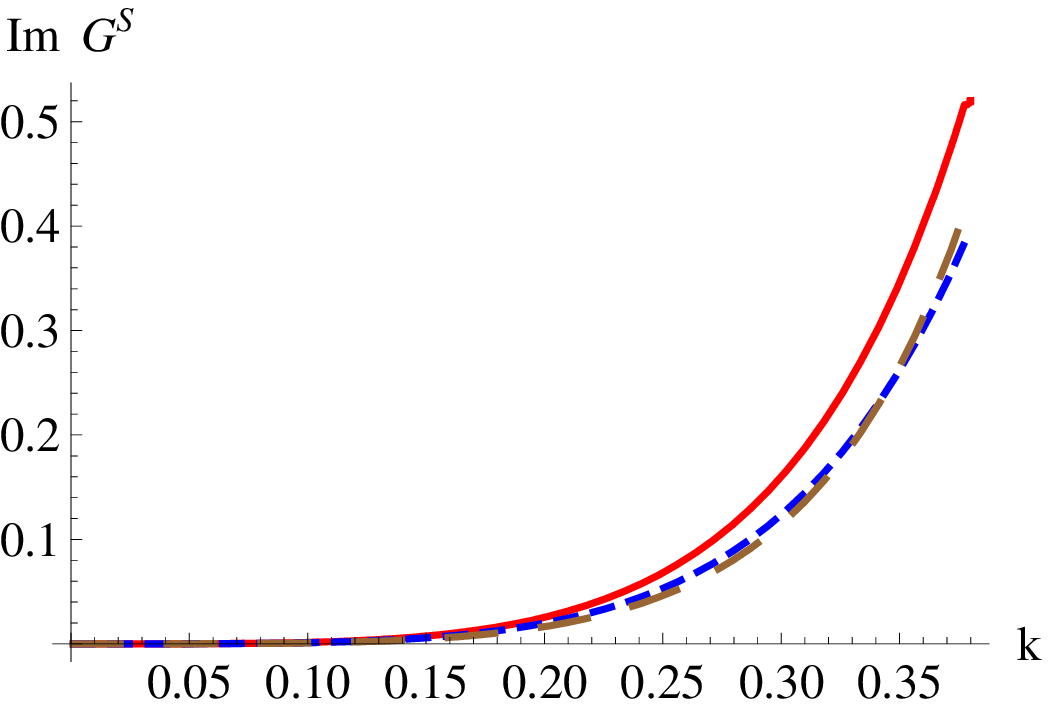,width=80mm} ~
\epsfig{file=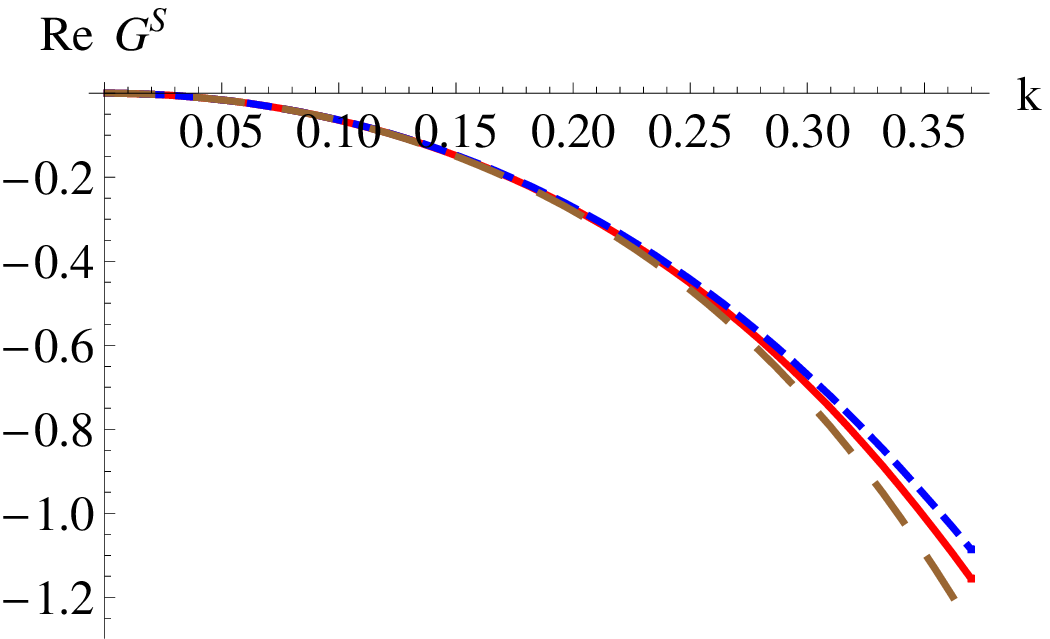,width=80mm}
\caption{\it Sound channel: top $k=0.4$; bottom $\omega=0.4$. Solid  line corresponds to the AdS/CFT correlator. Short dashes display the NS hydrodynamics while
long dashes show the IS hydrodynamics.}
\label{fig2}
}

\subsection{Life on the boundary: All order hydrodynamics}

The thermal field theory on the 4d boundary is defined by means of the generating functional 
\beq\label{Z}
Z[h]\,=\,\int D\phi\,e^{\,S_0[\phi]\ +\ \int d^4x\,h_{\mu\nu}\,T^{\mu\nu}}
\eeq
where $\phi$ collectively denotes all fields of the theory, $S_0$ is the flat metric action
and  $h_{\mu\nu}$ is an external perturbation of the Minkowski space. 

The expectation value of the energy-momentum tensor at non-vanishing external field $h_{\mu\nu}$ is
\beq\label{TFT}
\langle T^{\mu\nu}\rangle_{cl}^h\,=\,{\delta \ln Z\over \delta h}\,=\,
\langle T^{\mu\nu}\rangle_{cl}^{h=0}\,+\,
h_{\alpha\beta}\,\tilde G^{\alpha\beta\mu\nu}
\eeq
Within the linear response theory we keep terms linear in $h$ only. The correlators
$\tilde G^{\alpha\beta\mu\nu}$ differ from the retarded correlators 
$G^{\alpha\beta\mu\nu}$ by constant contact terms \cite{PSS}.

We use eq. (\ref{TFT}) to define  hydrodynamic variables.
Here $\langle T^{\mu\nu}\rangle_{cl}^{h=0}$ corresponds to the thermal equilibrium. 
The equilibrium energy density is 
\beq
\epsilon_0\,\equiv\,\langle T^{00}\rangle_{cl}^{h=0}
\eeq
The external perturbation
$h_{\mu\nu}$ shifts the theory from its thermal equilibrium.
The out-of- equilibrium energy density is
\beq
\epsilon\,\equiv\,\langle T^{00}\rangle_{cl}^{h}\,=\,\epsilon_0\,+\,h_{\alpha\beta}\,
\tilde G^{\alpha\beta00}
\eeq
We can also define fluid's three-velocity $v^i$ 
\beq
(\epsilon_0\,+\,P_0)\,v^i\,\equiv\,\langle T^{0i}\rangle_{cl}^{h}\,=\,h_{\alpha\beta}\,
\tilde G^{\alpha\beta0i}
\eeq
and fluid`s 4-velocity $u^\mu=(\sqrt{1+v^2},v)$ satisfying $u^2=-1$.

The action (\ref{Z}) is required to be invariant 
under the local Weyl transformation (see extensive discussion in Ref. \cite{BSSSR})
\beq\label{Weyl}
g_{\mu\nu}\,\rightarrow \,e^{-2\,\Omega(x,t)}\,g_{\mu\nu}\,.
\eeq
The invariance of the action implies that $T^{\mu\nu}$ and the  velocity field $u$ transform homogeneously
\beq
T^{\mu\nu}\,\rightarrow\,e^{6\,\Omega(x,t)}\,T^{\mu\nu}\,;
\ \ \ \ \ \ \ \  \ \  \ \ \ \ \ \ \ \ \ \ \ \ \ \ u^\mu\,\rightarrow \,e^{\Omega(x,t)}\,u^\mu\,.
\eeq
In our construction below we will be imposing the Weyl invariance. To this goal we will employ the 
Weyl tensor $C^{\lambda}_{\mu\nu \alpha}$ 
$$
C^\lambda_{\mu\nu\alpha}\,=\,R^\lambda_{\ \mu\nu\alpha}\,-\,{1\over 2}\,(g^\lambda_\nu\,R_{\mu\alpha}\,-\,
g^\lambda_\alpha\,R_{\mu\nu}\,-\,g_{\mu\nu}\,R^\lambda_{\ \alpha}\,+\,g_{\mu\alpha}\,R^\lambda_{\ \nu})
\,+\, {1\over 6}\,R\,(g^\lambda_\nu\,g_{\mu\alpha}\,-\,g^\lambda_\alpha\,g_{\mu\nu}),
$$
which is constructed to be invariant under this transformation. Here $R^\lambda_{\ \mu\nu\alpha}$, 
$R^\lambda_{\ \alpha}$ and $R$ stand for the Riemann, Ricci  tensors and the scalar curvature.

The hydro representation of the energy-momentum tensor 
\beq
\langle T^{\mu\nu}\rangle_{cl}^h\,=\,(\epsilon\,+\,P)\,u^\mu\,u^\nu\,+\,P\,g^{\mu\nu}\,+\,\Pi^{\langle\mu\nu\rangle}\,.
\eeq
Here for any tensor $\Pi^{\mu\nu}$ we define its traceless and symmetric component
(following the notations of Ref \cite{BSSSR})
\beq
\Pi^{\langle\mu\nu\rangle}\,=\,{1\over 2}\,\Delta^{\mu\alpha}\,\Delta^{\mu\beta}\,(\Pi_{\alpha\beta}\,+\,\Pi_{\beta\alpha})
\,-\,{1\over 3}\,\Delta^{\mu\nu}\,\Delta^{\alpha\beta}\,\Pi_{\alpha\beta}
\eeq
with the projector
\beq
\Delta^{\mu\nu}\,=\,g^{\mu\nu}\,+\,u^\mu\,u^\nu
\eeq
which is commonly introduced to ensure transversity of the tensor of dissipations $\Pi$:
\beq
u_\mu\,\Pi^{\mu\nu}\,=\,0\,.
\eeq
This transversity  is equivalent to the condition of no dissipation in the fluid`s rest frame.
The tracelessness of $T^{\mu\nu}$ implies that  $\Pi^{\langle\mu\nu\rangle}$ is also traceless 
\footnote{We ignore the Weyl anomaly since it is non-linear in the metric perturbations}. 
The metric $g$ is the full metric, namely the Minkowski metric perturbed by $h$.

The energy-momentum conservation leads to equations of motion for the fluid: 
\beq
\nabla_\mu\,\langle T^{\mu\nu}\rangle \,=\,0\,.
\eeq
Here $\nabla_\mu$ stands for covariant derivative with respect to the metric $g$.

The tensor  $\Pi^{\mu\nu}$ is considered to have  all order gradient expansion.
Within the linearized approximation discussed above, and constrained by the Lorentz and Weyl symmetries,
there are four independent structures (operators) one can write down
\footnote{To our understanding, there are no more structures one could possibly add 
to the expansion (\ref{Pi}). The only tensors with more than four Lorentz indices and
linear in $h$ are the ones obtained by applying covariant derivatives to the Weyl tensor. 
} 
\beq\label{Pi}
\Pi^{\mu\nu}\,=\,-\,2\,\eta\,\nabla^\mu\,u^\nu\,
%+\,\eta_2\,\nabla^\alpha\,\nabla^\mu\,\nabla^\nu\,u_\alpha\,
+\,
2\,\kappa\,u_\alpha\,u_\beta\,C^{\mu\alpha\nu\beta}\,+\,\rho\,
(u_\alpha\,\nabla_\beta\,+\,u_\beta\,\nabla_\alpha)\, C^{\mu\alpha\nu\beta}\,+\,
\xi\,\nabla_\alpha\,\nabla_\beta\,C^{\mu\alpha\nu\beta}\,.
\eeq
By representing $\Pi^{\mu\nu}$ in the form (\ref{Pi})
we essentially  postulate a constitutive relation
between $\langle T^{ij}\rangle $ and $v^i$.
This structure implies that the fluid can be perturbed either by inducing some velocity
perturbation or by shaking the metric. These perturbations are not fully independent
and  can be related by the equations of motion: the gravity perturbations create perturbations
of velocity (see Appendix B).

Each of the four transport coefficient 
functions $\eta$, $\kappa$, $\rho$ and $\xi$ are considered to be functions of the 
Lorentz scalar operators $\nabla^2$ and $(u\,\nabla)$
\beq
\eta\,=\,\eta[\nabla^2,(u\nabla)]\,;\ \ \ \ \ \ \ \ \
\kappa\,=\,\kappa[\nabla^2,(u\nabla)]\,;\ \ \ \ \ \ \ \ \
\rho\,=\,\rho[\nabla^2,(u\nabla)]\,;\ \ \ \ \ \ \ \ \
\xi\,=\,\xi[\nabla^2,(u\nabla)]\,;\ \ \ \ \ \ \ \ \
\eeq
In momentum space representation (adequate for our framework of linear approximation)  these functions
will depend on  $i\, \omega$ and $k^2$: $\nabla^2\,\rightarrow\,\omega^2\,-\,k^2$ and 
$(u\,\nabla)\,\rightarrow\,-\,i\,\omega$.

The first term generalizes the usual shear viscosity coefficient
$\eta_0$ defined at zero frequency and momentum. 
It also contains the relaxation time term of  second order hydrodynamics. 
%$\eta_2$ is another ``shear'' viscosity function. It is formaly introduced
%as a new transport coefficient of the third-order hydro. We will argue below that
%in fact the lowest order it can actually contribute is four. 
The other terms (GSFs) are due to  metric perturbations, absent in Minkowski space.
However, as was pointed out in Ref. \cite{BSSSR}, these terms contribute directly to two-point
functions of stress tensors, as computed from the bulk gravity side.  Also 
the ``$\kappa$'' term has been first introduced in Ref. \cite{BSSSR} 
\footnote{In \cite{BSSSR} $\kappa$ was introduced as 
constant.}.  

From the Minkowski perspective, the physical role of $\kappa$, $\rho$ and $\xi$ is not obvious.
It is well known that the correlators
of $T^{\mu\nu}$ contain not only  ``thermal'' physics but in addition get contaminated
by the vacuum or zero temperature contributions due to pair production (this is because the underlying microscopic
theory is a quantum field theory). 
However, a naive subtraction of $T=0$ contributions
leads to sign alternating results for imaginary parts of the correlators \cite{KS,Teaney,Yaffe} which cannot be identified
with true thermal spectral functions. This suggests a presence of interference terms between 
``vacuum'' and ``thermal'' amplitudes.

It is tempting to identify  the viscosity term with pure hydrodynamic (``thermal'') physics associated with the matter flow, 
and  the GSFs with the non-hydrodynamic or non-matter effects and the interference thereof. 
This conjecture is nicely supported by 
the $\xi$ term, which at first glance does not depend on the fluid`s velocity
and temperature  at all
\footnote{Up to non-linear terms it actually  coincides with the stress-energy tensor of the 
conformal gravity \cite{Man}.}. Consequently, when looking at the correlators, we  will find 
that in all three channels the contributions due to the $\xi$ term could be naturally 
identified with the vacuum ($T=0$) effects. 
The spectral functions computed from the viscosity terms only are positive definite,
as they should.

We would like to comment on the Weyl invariance and non-linear completions.
When introducing the all order (linearized) hydrodynamics (\ref{Pi}) we presented the tensor 
$\Pi^{\mu\nu}$ as transforming homogeneously
under the Weyl transformation (\ref{Weyl}). It is obvious, however, that $\Pi^{\mu\nu}$,
 as it appears in (\ref{Pi}), 
does not have this property. This is because, 
while the tensor $C^{\mu\alpha\nu\beta}$ 
is Weyl invariant, its derivatives are not. Furthermore, higher order derivatives
put as arguments of the transport coefficient functions also destroy the desired transformation properties.
The correct statement is that the Weyl invariance is recovered up to non-linear terms, 
which by themselves are of no interest to us  in this paper \footnote{ 
For the $\xi$ term with constant $\xi$ 
there exists a well known non-linear completion (see e.g. \cite{Man}): under the Weyl transformation
the tensor $\nabla_\alpha\,\nabla_\beta\,C^{\mu\alpha\nu\beta}\,-\,1/2\,C^{\mu\alpha\nu\beta}\,R_{\alpha\beta}$
transforms homogeneously.
}.

It is then a legitimate and interesting question to ask if for any higher order derivative term there exists a non-linear
completion needed to restore the right transformation property under the Weyl transformation. 
Can it happen that some  of the higher order derivatives both in the viscosity term and the GSF
terms cannot be completed to meet the requirement of the Weyl invariance and should be forbidden (similarly to the fate
of the bulk viscosity term)?  The answer is negative and
there is no additional selection principle based on the Weyl symmetry.
For any number of derivatives there exist a non-linear completion with
the formal construction given in Ref. \cite{Log}. 
It is based on the fact that, instead of the covariant derivative $\nabla^\mu$, one can introduce 
an even longer derivative $D^\mu$ involving the Weyl connection constructed from the field $u$ itself. 
Any number of these derivatives acting on $C^{\mu\  \nu}_{\ \alpha\ \beta}$
 leaves  a Weyl-invariant result. 
This procedure generates non-linear terms, which  are of no  interest to us in  this paper. 
For our purposes it is sufficient to know about their existence. 
Note, however, that the procedure of Ref. \cite{Log} can be used to reconstruct these non-linear terms%Most of the original applications of hydrodynamics to RHIC physics were focused mostly on flows characterizing
%global expansions of the plasma fireball (such as radial and eliptic flows).  Typical gradients
 %associated
%with these expansions are $1/R$ (nucleus radius),
 %which leads to a dimentionless momenta $1/2\,\pi\,T\,R\,\simeq $ ???????
%It is important to stress that the global expansions are governed by non-linear hydrodynamics.
%The same is true for the Bjorken process considered by us in Ref. \cite{LS}. In other words, one cannot
%use any linear approximation in order to reliably describe the fireball explosion. Thus one might suspect
%that there are no realistic applications of the linearized hydrodynamics. 
%We would like to argue that this is far being true.

% why those are needed?
%QCD sum rules. resonance + continuum

from the higher order linear terms discussed below. That would certainly provide more insight on hydrodynamics
at order three and higher.

If we were not to impose the Weyl invariance, we would introduce another shear viscosity term in the expansion
(\ref{Pi}),
$$
\eta_2\,\nabla^\nu\,\nabla^\mu\,\nabla^\alpha\,u_\alpha
$$
This term would normally contribute to the sound channel starting from  third order hydrodynamics. 
We would like to argue that this term is in fact forbidden by  Weyl invariance. As was explained above, in order to 
comply with  Weyl invariance the correct prescription is to use long derivatives $D^\alpha$ instead of 
$\nabla^\alpha$. 
However, the long derivative $D^\alpha$ has the property $D^\alpha\,u_\alpha=0$, 
which eliminates the $\eta_2$ term.

The hydro ansatz (\ref{Pi}) can be probed by small gravity perturbations. Using     
linear response theory we can then compute the retarded correlators in the three
symmetry channels (the computation is presented in Appendix B). 

\noindent $\bullet$ The scalar:
\beq\label{GT}
G^T(k,w)\,=\,-\,i\,\omega\,\eta\,-\, \kappa\,{1\over 2}\,(w^2 \,+\,k^2)\,-\,
\rho\,{i\,\omega\over 2}\,(w^2 \,- \,k^2)\,+\,
\xi\,{1\over 4}\,(\omega^2\,-\,k^2)^2
\eeq
$\bullet$ The shear:
\beq\label{GD}
G^D(k,w)\,=\,(\epsilon\,+\,P)\ 
\frac{\bar\eta\,k^2\,-\,i\bar\kappa\,\omega\,k^2/2\,-\,\bar\rho\,k^2\,(k^2\,-\,2\,\omega^2)/4\,+
\,i\,\bar\xi\,\omega\,k^2\,(\omega^2\,-\,k^2)/4}{-i\,\omega\,+\,\bar\eta\,k^2}
\eeq
$\bullet$ The sound:
\beq\label{GS}
G^S(k,w)\,=\,(\epsilon\,+\,P)\ \frac{k^2\,-\,4\,i\,\bar\eta\,\omega\,k^2\,
-\,2\,\bar\kappa\,\omega^2\,k^2\,-\,2\,i\,\bar\rho\,\omega^3\,k^2\,+\,\bar\xi\,\omega^4\,k^2}
{k^2\,-\,3\,\omega^2\,-\,4\,i\,\bar\eta\,\omega\,k^2}
\eeq
with 
\beq
\bar \eta\,\equiv\, \eta/(\epsilon\,+\,P)\,;\ \ \ \ \ \ \ \ \
\bar \kappa\,\equiv\, \kappa/(\epsilon\,+\,P)\,;\ \ \ \ \ \ \ \ \
\bar \rho\,\equiv\, \rho/(\epsilon\,+\,P)\,;\ \ \ \ \ \ \ \ \
\bar \xi\,\equiv\, \xi/(\epsilon\,+\,P)\,.
\eeq
%and 
%\beq
%\eta\ \equiv\ \eta_1\,; \ \ \ \ \ \ \ \ \ \ \ \ \  \eta_{12}\,\equiv\,\eta_1\,+\,\eta_2\,(\omega^2\,-\,2\,k^2)
%\eeq
Note that when $k=0$ the SO(3) symmetry of the space is restored. Modulo trivial rescaling we do indeed observe
that the three correlators $G^T$, $G^D$, and $G^S$ all coincide:
$$
G^T|_{k\rightarrow 0}\,=\,-\,{\om^2\over k^2}\,G^D|_{k\rightarrow 0}\,=\,-\,
3/4\,{\om^2\over k^2}\,G^S|_{k\rightarrow 0}\,-\,(\epsilon\,+\,P)/4
$$
%, provided
%\beq
%\eta_2(k\rightarrow 0,w)\,=\,0
%\eeq
%This means $\eta_2$ even if not vanishing identically is proportional to $k^2$, and thus consitute a correction to $\eta_1$
%at minimum order four (fifth order hydrodynamics). 

At large frequencies $w\gg 1$, the temperature effects should be negligible and
 the correlators $G$ are expected to coincide with the correlators computed in the vacuum:
\begin{eqnarray}\label{Gas}
&&G^T(\om,k)_{T=0}\,=\,(\epsilon\,+\,P)\ (\omega^2\,-\,k^2)^2\,\ln (k^2\,-\,\om^2) \,; \nonumber \\
&&G^D(\om,k)_{T=0}\, =\,-\,(\epsilon\,+\,P)\ k^2\,(\omega^2\,-\,k^2)\,\ln (k^2\,-\,\om^2) \,; \nonumber \\
&&G^S(\om,k)_{T=0}\, =\,-\,(\epsilon\,+\,P)\ (4/3)\ k^2 (\omega^2\,-\,k^2)\,\ln (k^2\,-\,\om^2)\,. 
\end{eqnarray}
The asymptotics (\ref{Gas}) is indeed observed in the correlators computed from the bulk gravity (see previous section).
What is interesting to note that the behavior (\ref{Gas}) is naturally identified with the $\xi$ terms in the correlators,
suggesting $\xi\,\sim\,\ln (k^2\,-\,\om^2)$ at asymptotically large  $\om$. It is then tempting to identify the $\xi$ terms
as responsible for 
the  contribution to the correlators of the non-hydro pair creation effects, 
while the $\kappa$ and $\rho$ terms could be regarded as interference contributions
between the ``vacuum'' and ``hydro'' physics. Within such interpretation it is natural to identify $\eta$ as purely
hydrodynamical  effects associated with the matter flow. 
Thus if one is interested in pure thermal/hydrodynamic correlators, one first has to determine $\eta$ as 
functions of momenta and then compute the correlators with the GSFs set to zero.

Despite this nice interpretation of $\xi$ as the pure ``vacuum''  term, all GSF terms in fact fully 
mix when considered as  functions of momenta. If we consider $(\om\rightarrow 0,\,k\rightarrow\infty)$  asymptotics, all
correlators tend to behave proportional to $k^4\,\ln k^2$. From this behavior we can learn about the asymptotic
behavior of the GSFs themselves
\beq
\kappa\,\sim\,k^2\,\ln k^2\,, \ \ \ \ \  \ \ \ \ \ \ 
\rho\,\sim\,\sqrt{k^2}\,\ln k^2\,,\ \ \ \ \ \ \ \ \ \ \ 
\xi\,\sim\,\ln k^2\,.
\eeq

\section{When the bulk meets the boundary: Results}

There should be  one to one correspondence between linearized $T^{\mu\nu}$ and the full set of its correlators.
Our program is to equate the expressions (\ref{GT},\ref{GD},\ref{GS}) for the correlators to the correlators 
computed from the bulk gravity. The goal is to invert these equations in order to  determine the
four transport coefficient functions. We have got an apparent problem as
we end up having only three equations for four unknown functions.
This system does not seem to have a unique solution. Despite our failure to simultaneously determine
all transport coefficient functions, we are able to extract them perturbatively in the long-wave limit  approximation. 
%For the rest of this paper we assume $\eta_2=0$.

In the near-longwave limit all of the coefficient functions are expandable in  power series
\footnote{We belive this expansion has a finite radius of convergence, The radius of convergence 
is given by the first singularity, which coincides with 
the first quasinormal mode of the scalar channel.}
\begin{eqnarray}
&&\eta\,=\, \eta_0(1+i\eta_{0,1}\,\omega + \eta_{2,0}\,k^2 + \eta_{0,2}\,w^2 + i\,\eta_{2,1}\,
\omega\,k^2
\,+\,i\,\eta_{0,3}\,\omega^3 + \eta_{4,0}\,k^4 + \eta_{2,2}\,\omega^2\,k^2 + \eta_{0,4}\,\omega^4+\cdots);
 \nonumber \\
&&\kappa\,=\, \kappa_0\,(1\,+\,i\,\kappa_{0,1}\,\omega\,+\,\kappa_{2,0}\,k^2\,+\,\kappa_{0,2}\,w^2\,+i\,\kappa_{2,1}\,\omega\,k^2\,+
\,i\,\kappa_{0,3}\,\omega^3\,+\,\cdots)\,;
\nonumber \\ 
&&\rho\,=\, \rho_0\,(1\,+\,i\,\rho_{0,1}\,\omega\,+\,\rho_{2,0}\,k^2\,+\,\rho_{0,2}\,w^2\,+\,\cdots)\nonumber \\ 
&&\xi\,=\, \xi_0\,(1\,+\,i\,\xi_{0,1}\,\omega\,+\,\cdots)
\label{exp}
\end{eqnarray}
Here we explicitly list all terms up to fifth order. % (ignoring $\eta_2$).
The third order coefficients are determined (practically all) analytically. 
The other coefficients are extracted numerically. We achieved a good
accuracy with the forth order  coefficients while the rest have large errors. 
\begin{eqnarray}
&&\eta_0\,=\,(\epsilon\,+\,P)/2;\ \ \ \ \ \ \ \ \ \ \ \ 
\tau\,\equiv\,\eta_{0,1}\,=\,{2\,-\,\ln 2}\,;\ \ \ \ \ \ \ \ \ \ \ \ \ 
\eta_{2,0}\,=\,-\,{1/2};\nonumber  \\ 
&&\kappa_0\,=\,{2\,\eta_0}\,;\ \ \ \ \ \ \ \ \ \ \ \ \ \ \ \ \ \ \ \
\kappa_{0,1}\,=\,{{5/2}\,-\,2\,\ln 2}\,;\ \ \ \ \ \ \ \ \ \ \ \ \ \
\rho_0\,=\,{4\,\eta_0}
\end{eqnarray}
The viscosity $\eta_0$ is of course just (\ref{bound}). 
The coefficient $\eta_{0,1}$ is the relaxation time,
 which within the AdS/CFT approach was first addressed 
in Ref. \cite{janik}. It was correctly
determined   in Ref. \cite{BSSSR, Minwalla} and later in \cite{Jap}. 
In \cite{BSSSR} it was found by looking at the first correction
to  speed of sound. $\eta_{0,1}$ can be consistently deduced from any of the three correlators.
$\kappa_0$ was found also in \cite{BSSSR} by matching the $k^2$ term in $G^T$.
 Independently and consistently, it can be
also found from the shear and
sound channels (the $\om\,k^2$ term in the numerator of $G^D$ and the
$w^2\,k^2$ term in the numerator of the function $G^S$).  

The coefficient $\eta_{2,0}$  appears at  third order hydro, 
which was left beyond the scopes of \cite{BSSSR}. However, this coefficient
could be easily read off from the analysis of Ref. \cite{BSSSR}, in particular, from the $k^4$ correction to
the diffusive pole in the shear channel. The result is consistent with the $k^4$ term in the numerator of $G^D$.
The coefficient $\rho_0$  is deduced from the $\omega=0$ limit of
the function $G^D$. Finally we analytically extracted the coefficient $\kappa_{0,1}$. This comes from  matching the 
coefficients of the $\omega\,k^2$ in the scalar channel.

The remaining coefficients were found numerically. Let consider the coefficient $\eta_{0,2}$
as an example of our numerical procedure.  We were able to get a very accurate fit of the coefficient in front 
of the $\omega^3$ term in the expansion of the correlator $G^T$. This coefficient is then trivially
related to $\eta_{0,2}$ and $\kappa_0$, $\kappa_{0,1}$, $\rho_0$, the latter being all previously determined.
The result is 
\beq\label{eta02}
\eta_{0,2}\,\simeq\,-\, 1.379\,\pm\,0.001\,\simeq\,-\, {3\over 2}\,+\,{\ln^2 2\over 4}
\eeq
where the last expression is our guess for the analytic expression. 
The error in eq. (\ref{eta02}), as well as other errors quoted below, reflect our confidence
in the results provided.

Despite the fact that we were not able to find a method to extract four unknown coefficient functions from three
equations, there seems to be a recurrent procedure,
 which make this task possible, at least perturbatively near
the long wave limit. The coefficient $\kappa_{2,0}$ can be obtained from the $\omega=0$ limit of the sound correlator
$G^S$. Once this one is known, the  $\omega=0$ limit of $G^T$ reveals the coefficient $\xi_0$, etc.

Below we present our numerical results.\\
4th order hydro  
\beq
\eta_{2,1}\,=\,-\,2.275\,\pm\,0.005\,;\ \ \ \ \ \ \ \ \ \ \ \ \ \ \ \ \ \ \  \
\eta_{0,3}\,=\,-\,0.082\,\pm\,0.003\,
\eeq
5th order hydro
\beq
\eta_{4,0}\,=\,0.565\,\pm\,0.005\,;\ \ \ \ \ \ \ \ \ \ \ \ \ \ \ \ 
\eta_{0,4}\,=\,2.9\,\pm\,0.1\,;\ \ \ \ \ \ \ \ \  \ \ \ \ \ \eta_{2,2}\,=\,1.1\,\pm\,0.2\,;    
\eeq
The GSF's coefficients
\begin{eqnarray}
&&
\kappa_{2,0}\,=\,-\,1.6\,\pm\,0.05\,;\ \ \ \ \ \ \ \ \ \ \ \ \ \ \ \ \ \ \ \ \ \ \  
\kappa_{0,2}\,=\,0.04\,\pm\,0.01\,;\nonumber \\
&&\kappa_{0,3}\,=\,-\,1.95\,\pm\,0.05\,;\ \ \ \ \ \ \ \ \ \  \ \ \ \ \ \ \ \ \ \ \ \
\kappa_{2,1}\,=\,-\,1.6\,\pm\,0.2\,;
\nonumber \\
&&
\rho_{0,1}\,=\,0.92\,\pm\,0.01\,;\ \ \ \ \ \ \ \ \ \ \ \ \ \ \ \ \ \ \ \ \ \ \ \ \ 
\rho_{0,2}\,=\,-\,0.68\,\pm\,0.04\,;\ \ \ \ \ \ \ \
\rho_{2,0}\,=\,-\,0.755\,\pm\,0.005\,;\nonumber \\
&&\xi_{0}\,=\,-\,2.6\,\pm\,0.1\,;\ \ \ \ \ \ \ \ \ \ \ \ \ \ \ \ \ \ \ \ \ \ \ \ \ \ \ 
\xi_{0,1}\,=\,-1.1\,\pm\,0.2\,; 
\end{eqnarray}

To summarize our knowledge of viscosity function $\eta$, we plot it  and compare  to the IS one (Fig. \ref{fig3}).
\FIGURE{\epsfig{file=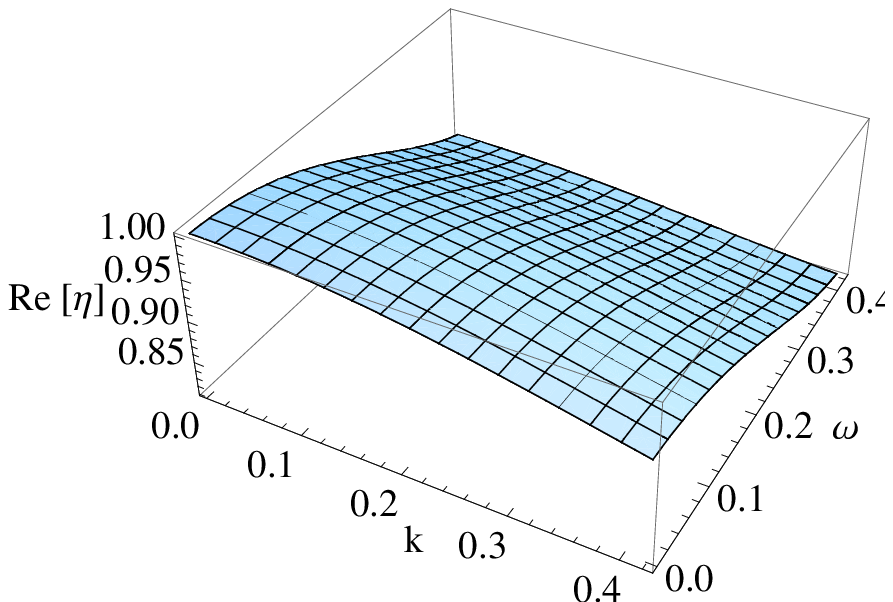,width=75mm} ~
\epsfig{file=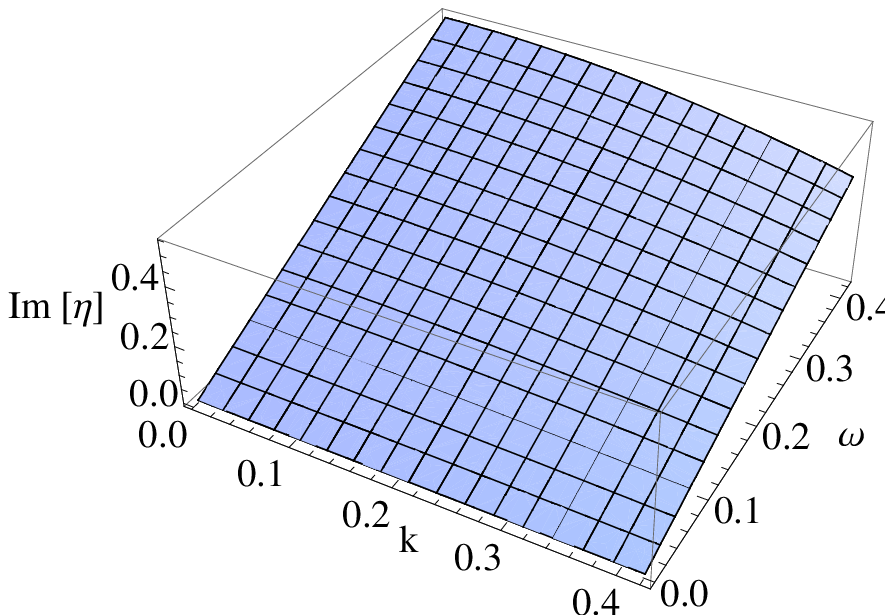,width=75mm}
\epsfig{file=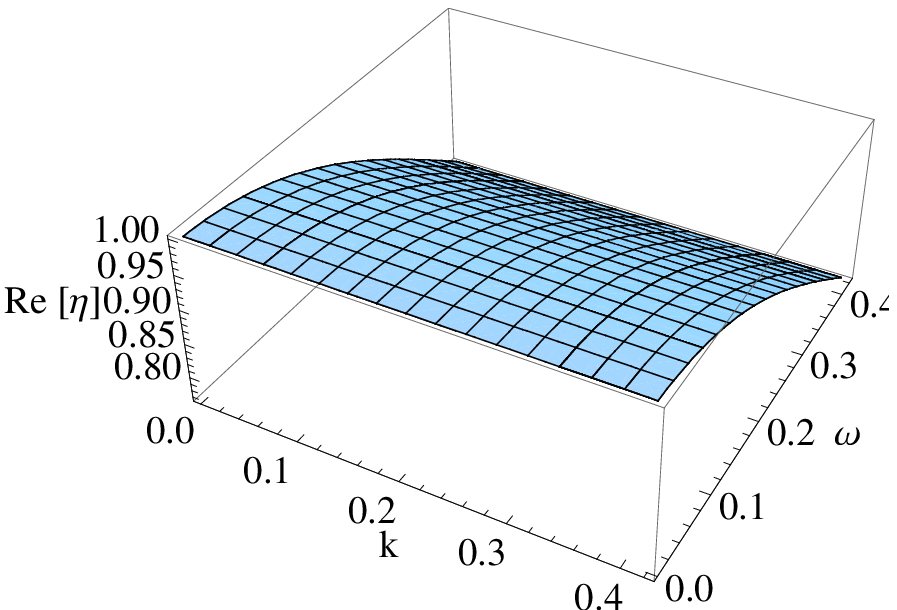,width=75mm} ~
\epsfig{file=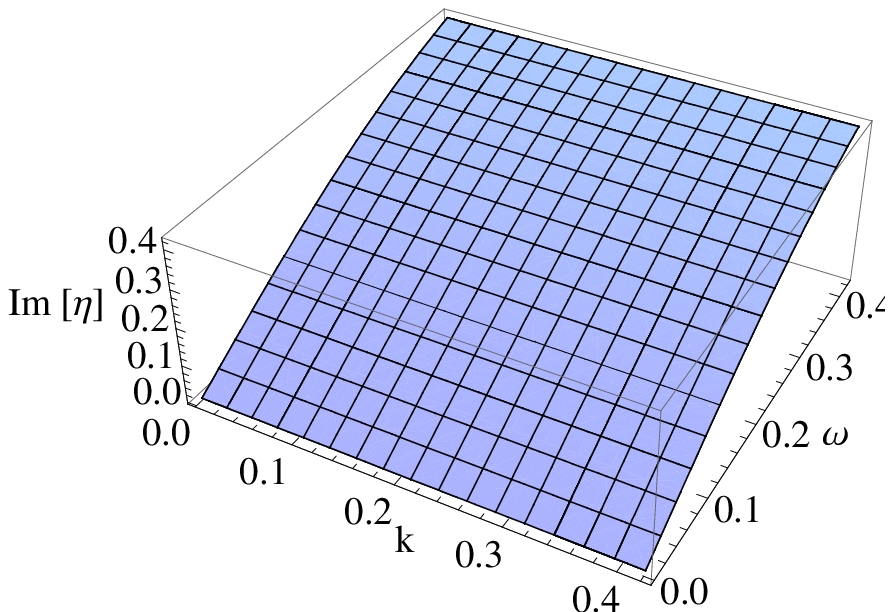,width=75mm}
\caption{\it Viscosity function (divided by $\eta_0$): top AdS/CFT; bottom IS.}
\label{fig3}
}
The NS value is, of course, $\eta=\eta_0$. 
For $\om,\,k \le 0.4$ we can expect up to 15\% correction due to 
momenta-dependence of the viscosity function.

\subsection*{On quasinormal modes and analytic structure of the 
viscosity function}

The quasinormal modes are poles of the retarded correlators. 
They have been analyzed in all three channels  in \cite{KSqnm}.
We would like to argue that  entire information about quasinormal modes 
is coded in the viscosity function $\eta$,
while the GSF do not have any poles. If this were not true, we would observe appearance of 
identical quasinormal modes in all three
channels, which is not the case at least for a number of low lying modes. 
\beq
\eta(k^2,w)\,=\,\sum_{n=0}^{\infty} \,{\eta_n(k^2,w)\over\omega\,-\,\omega_n(k^2)}
\eeq
We further argue that $\omega_n$ coincide with the quasinormal modes of the scalar channel 
(poles of $G^T$), 
which have been analyzed in the past (see the table below). 
At $k^2=0$ they can be computed quasiclassically for large $n$ (in fact quasiclassics works well down to $n=2$)
\cite{QNM}
\beq
\omega_n(k^2=0)\,\simeq\,\omega_0\,+\,n\,(\pm\,1\,-\,i)
\eeq
No analytical expression for non-zero $k$ is  known.

The quasinormal modes of the shear and sound channels are obtained from the following dispersion relations.
\beq\label{dr}
-\,i\,\omega\,+\,\eta(k^2,\omega)\,k^2\,=\,0; \ \ \ \ \ \ \ \ \ \ \ \  \ \ \ \ \ \ \ 
-\,3\,\omega^2\,+\,k^2\,-\,4\,i\,\eta(k^2,\omega)\,\omega\,k^2\,=\,0
\eeq
As well known,
 these dispersion relations admit hydrodynamic modes as  lowest modes in the spectrum.
Higher modes will appear as  distorted  spectrum $\omega_n$. 
Furthermore, the higher the mode the less
distortion should be present. In other words, the spectra of all three channels will become degenerate
for high modes. This tendency is clearly observed in the following table copied from Ref. \cite{KSqnm} 
($k^2\,=\,1$)
\begin{center}
\begin{tabular}{|c||c|c||c|c||c|c|}
\hline
{}  & 
\multicolumn{2}{c||}{Scalar channel} & 
\multicolumn{2}{c||}{Shear channel} & 
\multicolumn{2}{c|}{Sound channel} \\ \cline{2-7}
$n$ & $\Re e\ \omega_n$ & $\Im m\ \omega_n$ & $\Re e\ \omega_n$ & $\Im m \ \omega_n$ & $\Re e\ \omega_n$ & $\Im m\ \omega_n$\\
\hline
$1$ & $\pm$1.954331 & $-$1.267327 & $\pm$1.759116 & $-$1.291594 & $\pm$1.733511 & $-$1.343008 \\
$2$ & $\pm$2.880263 & $-$2.297957 & $\pm$2.733081 & $-$2.330405 & $\pm$2.705540 & $-$2.357062 \\
$3$ & $\pm$3.836632 & $-$3.314907 & $\pm$3.715933 & $-$3.345343 & $\pm$3.689392 & $-$3.363863 \\
$4$ & $\pm$4.807392 & $-$4.325871 & $\pm$4.703643 & $-$4.353487 & $\pm$4.678736 & $-$4.367981 \\
$5$ & $\pm$5.786182 & $-$5.333622 & $\pm$5.694472 & $-$5.358205 & $\pm$5.671091 & $-$5.370784 \\
\hline
\end{tabular}
\end{center}

Finally we would like to note that from the behavior of the sound dispersion curve  \cite{KSqnm}
one can deduce the following asymptotic behavior of the viscosity function
\beq\label{asym}
\eta(k^2\sim \om^2\,\rightarrow\, \infty)\,\rightarrow\,{i\over 2\,\omega}\,,
\eeq
which supports our understanding that it is a falling function at large momenta.

\section{Model for improved causal hydrodynamics}

Though in this paper we do not pursuit any practical applications,  we
would like to propose an improved and causal hydrodynamics for future use by hydro practitioners.

While we were not able to achieve our prime goal, of deducing the viscosity function in full range of frequency and momentum,
we were  able to get several new coefficients for the small momenta expansion. Below we present a resummation scheme
similar to IS, which is an ansatz aimed at providing a good model for the entire viscosity function. The model is constructed
with the requirement of causality built in.

Causality implies that the imaginary part of the poles is always negative and the function vanishes at infinite frequencies.
This is equivalent to the validity of the dispersion relation:
\beq
\eta(k^2,\om)\,=\,\int \,{d\om^\prime\over 2\,\pi\,i}\,{\Re e\,\eta(k^2,\om^\prime)\over \om^\prime\,-\,\om}
\eeq
In addition, in order to relate the viscosity function to the thermal spectral functions,
we require that both real and imaginary parts of it remain positive for
all  values of momentum and frequency. 

Similarly to the IS model,
 we take a Pade-like resummation ansatz which reproduces all
 low momentum coefficients in the expansion.
\beq
\eta_{model\,^1}\,=\,\eta_0\,\sum_{i=1}^3\,{d_i\over a_i\,+\,b_i\,k^2\,-\,i\,w}
\eeq
This ansatz has three pure imaginary poles and it reproduces exactly eight first coefficients in the expansion (\ref{exp}).  
$$
d_1=0.736\,,\ \ \ \ \ a_1=0.72731\,,\ \ \ \ \ \ b_1=0.3263 \ \ \ \ \ \ 
d_2=2.1\, , \ \ \ \ \ a_2=0.10618\,, \ \ \ \ \ \ b_2=0.3042\,,
$$
$$
d_3=-2.1016\,, \ \ \ \ \ \ a_3=0.10620\,,\ \ \ \ \ \ \ b_3=0.3038\,.
$$
The resummed viscosity function is plotted in Fig. \ref{fig4}.
\FIGURE{\epsfig{file=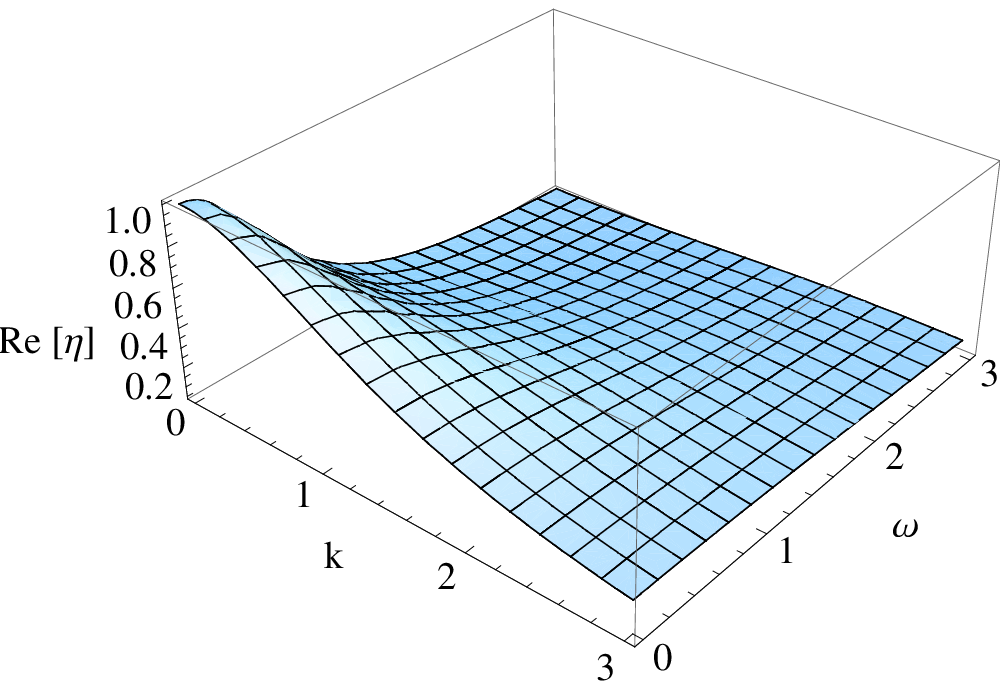,width=80mm} ~
\epsfig{file=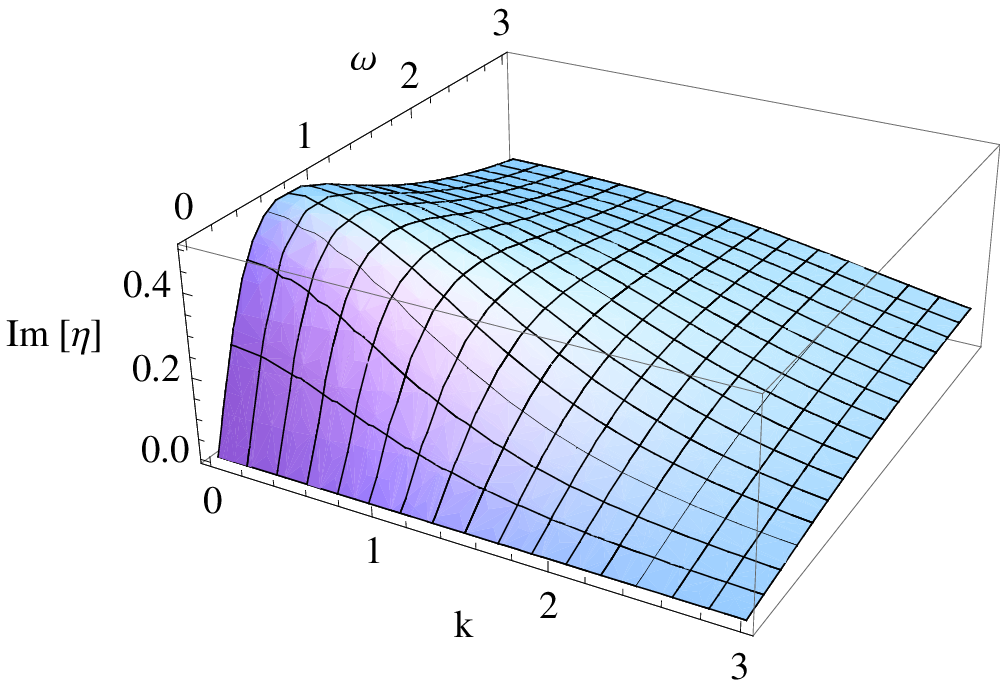,width=80mm}
\caption{\it Viscosity function (divided by $\eta_0$): the model}
\label{fig4}
}
This model could be further improved by accounting for the asymptotic behavior (\ref{asym})
as well as for information about   quasinormal modes of the scalar channel.
The second and third poles practically cancel each other. 
 Despite the fact that it does not accurately
reproduce the expansion, it  turns out to be a  very good approximation
to retain only one pole, similarly to IS but with three-momentum dependence. 
\beq
\eta_{model\,^2}\,=\,{\eta_0\over 1\,-\,\eta_{2,0}\,k^2\,-\,i\,w\,\eta_{0,1}}
\eeq
Within about 10\% accuracy (and in some regions  with much better one) 
the second model  is equivalent to the first one. Since the entire effect
of momenta-dependence is not expected to be very large,
 the second model should be more than sufficient for any phenomenological
applications. We note that the group velocity for the sound mode computed
within this model is always smaller than one, confirming  causality of the model.

The viscosity function can be Fourier transformed into the memory function 
\beq
D(x,t)\,=\,\int d\om\,d^3k\,e^{-i\,\om\,t\,+\,i\,k\,x}\ \eta(k^2,\om)
\eeq
which leads to the following expression for the dissipation tensor $\Pi$:
\beq
\Pi^{\mu\nu}\,=\,-\,2\,\int_0^t\,dt^\prime\,\int d^3 x^\prime \,D(x-x^\prime,t-t^\prime)\ \nabla^{\prime\,\mu}\,
u^\nu(x^\prime,t^\prime)
\eeq
%The model constracted above do satisfy this dispertion relation
Performing the Fourier transform explicitly we obtain
\beq
D_{model\,^2}(x,t)\,=\,\int d\om\,d^3k\,e^{-i\,\om\,t\,+\,i\,k\,x}\ \eta_{model\,^2}(k^2,\om)\,=\,
{1\over 2\,\sqrt{2}}\,
{\eta_0\over\eta_{0,1}}\ \left({-\,\eta_{0,1}\over \eta_{2,0}\,t}\right)^{3/2}\ e^{\,-\,t\,/\,\eta_{0,1}}\ 
e^{\,x^2\, \eta_{0,1}\,/\,(\eta_{2,0}\,t)}
\eeq
We remind the reader that $\eta_{2,0}$ is negative.

\section{Summary and Discussion}

In this paper we initiated a study of all order velocity gradient expansion of linearized relativistic hydrodynamics
near equilibrium. The research was carried out within the ${\cal N}=4$ SYM theory at large $N_c$. More specifically, 
we parameterized the energy-momentum tensor of the theory in terms of four momenta-dependent functions. 
These functions  generalize the notion of the usual constant transport coefficients, such as  viscosity,
into momenta dependent ones. We then attempted to determine all four functions based on the information on 
retarded correlators of two stress tensors. The latter were computed via the AdS/CFT prescription for computing
retarded correlators from bulk gravity waves.
 
Out of four transport coefficient functions, which we introduced in (\ref{Pi}), $\eta$ 
appears as a coefficient of the operator constructed from velocity gradients and
is a  generalization of shear viscosity. The remaining three coefficients (GSFs) 
arise as coefficients of 4d metric perturbations and appear in front of three operators involving the curvature Weyl
tensor. 

In this paper we extended the previous knowledge of hydrodynamic 
transport coefficients  at first and second order  to some higher order coefficients. We were able to find only those,
 which contribute to linearized hydrodynamics. We gave analytic values for two  coefficients of the 
third order hydro. We provided very accurate numerical estimates
for two coefficients of the forth order and one of the fifth. In addition, we introduced and determined several new coefficients
associated with the GSFs.

To illustrate the effect of the higher order terms in the viscosity function, we compute the sound 
dispersion curve by solving (perturbatively) eq. (\ref{dr}) to the order $k^6$:
\beq
\om_{AdS}\,=\,\pm\,{k\over \sqrt{3}}\,
\left(1\,+\,\left({1\over 2}\,-\,{\ln[2]\over 3}\right)\,k^2\,-\,0.088\,k^4\right)\,-
\,i\,{k^2\over 3}\,\left(1\,-\,{k^2\over 12}\,
(4\,-\,8\,\ln2\,+\,\ln^2 2)\,-\,0.15\,k^4\right)
\eeq
As we have
already emphasized in Ref. \cite{LS}, the sound width  gets negative corrections from  higher order terms.
This is in sharp contrast to the IS model, which leads to a qualitatively opposite effect, with the correction being
positive
\beq
\om_{IS}\,=\,\pm\,{k\over \sqrt{3}}\,
\left(1\,+\,\left({1\over 2}\,-\,{\ln[2]\over 3}\right)\,k^2\right)\,-\,i\,{k^2\over 3}\,
\left(1\,+\,{k^2\over 3}\,\ln 2\,(2\,-\,\ln 2)\right)
\eeq
Based on the new information about higher order terms in the expansion of the viscosity function,
we have proposed an improved causal IS-like (single pole) hydrodynamics, which we hope can be used
by hydro practitioners. Compared to IS, this  model  emphasizes the importance of the space momentum dependence
of the viscosity function. It leads to qualitatively different predictions, as seen from the width of the sound
pole. On the basis of this example, we cautiously suggest that the results based on the 
IS theory might be in fact less reliable than it was previously thought.
We also propose to exploit our improved  model for non-linear phenomena (such as Bjorken expansion and elliptic
flow) even though
such applications have no  theoretical justification.

Admittedly, the problem we had set up was not yet fully solved by the present paper.
Using a perturbative procedure, we  found several new higher order (constant) 
transport coefficients, either analytically or numerically. Nevertheless,
a  generic problem remains to be solved: the 
correlators we used as our input seemed not to be sufficient to determine
the four transport coefficient functions parameterizing all relevant kinematic structures. 
While it is  possible to follow  the  iterative approach used by us to 
 determine more coefficients in the expansion of the transport functions, we have no  proof
that this procedure will actually work at $all$ higher orders with  unique results. 
It is not excluded that some additional inputs (apart of the correlators)
are required in order to solve the problem in full.

As an  alternative approach to the problem, one may switch
 from solving the bulk equations for gravity waves -- the basis for computing  the correlators
 -- to a membrane paradigm-type approach based on  vibrations and translations of the horizon, as it is done in
 Ref. \cite{Minwalla} and the follow on papers \cite{Min2,Heller}. This approach 
provides  a quite general procedure to derive the next order derivative terms for  boundary
  hydrodynamics. In this approach,
  % one solves the bulk GR perturbatively 
%and then reads off
 the boundary energy-momentum tensor with appropriate gradient corrections is obtained
 through the usual holographic renormalization
procedure, % The bulk perturbation theory  is mapped onto the velocity gradient expansion on the boundary.  
with the bulk solution reflecting perturbation of the near-horizon ``membrane".
If the boundary metric is not taken as flat Minkowski (as it was done in \cite{Minwalla}), but rather as a
slightly perturbed one, the method of \cite{Minwalla} would reveal the GSFs alongside the viscosity function.
Furthermore, the approach  of \cite{Minwalla} has a potential to determine not only linear but also non-linear terms,
the latter being beyond the scope of our present paper. Here we obviously mean third and higher order hydrodynamics.
We have not pursued this direction, but believe it is worth studying it as it is important to 
learn about the gradient structure as a way to understand the non-equilibrium effects in  plasma.

An important general problem  is a separation between the hydrodynamic  
(thermal) physics associated with the matter flow and the vacuum (zero temperature) effects
associated with the pair production, as both contribute to the retarded correlators. 
We hope that we proposed the right approach to it, by identifying the
 different roles played by the viscosity function and the GSFs. While the
former is purely ``hydrodynamical'', the latter includes  non-thermal physics and  interference. This
separation of roles is very plausible, supported by the results at hand, 
but it was not proven by us in general.  We have argued in the text that the
pole structure of the correlators is entirely included in the viscosity function, while the GSFs 
have no poles. The overall role of the GSFs is somewhat unclear. On the one hand, they are 
formally introduced in (\ref{Pi}) as a response of the fluid to external gravitational shakings. 
On the other hand, from the analysis of the correlators we identify the GSFs as being responsible 
for flat space non-hydrodynamic (non-thermal) effects associated with pair production of the underlying
microscopic field theory. The metric perturbations  in effect mimic  non-hydro physics.

Are the GSFs relevant for RHIC experiment? We believe the answer is ``no''. We had to deal with them only
because the correlators used for our analysis contain both  thermal physics and  vacuum effects (such as pair
creation). One, of course, could propose another type of experiment, in which plasma would be exposed to a real
gravitational wave. In this type of experiment, the GSFs would determine the physical response of the fluid.

The so-called contact (or Schwinger) terms are QFT phenomena originating in UV. One could suspect that the GSFs originate
from those. However,
if this were the case, the only effect
they would produce is to shift the correlators by finite order polynomials in momenta. 
This is not the case, however. From the explicit expressions for the correlators (\ref{GD}, \ref{GS}) one can see that the 
numerator terms involving the GSFs cannot cancel the corresponding poles.  Thus they include more physics than
just  the contact terms part of which is not coming from the UV.

%It would be interesting to understand the physical meaning of the second shear viscosity $\eta_2$, provided
%it is not zero.  We have proven in the text that the lowest order it can contribute is five, meaning that at least two
%first coefficients of $\eta_2$`s small momenta expansion are zero. It is quite possible that $\eta_2$
%vanishes identically, but we have not found any general (symmetry?) argument which would impose that. 
%We regard the issue of $\eta_2$ existence as an open question. 

Last but not least,
it remains to be seen how relevant  the effects of momenta-dependent viscosity are 
 for realistic applications of relativistic hydrodynamics to heavy ion collisions. Our previous paper
\cite{LS} argues that they might be quite substantial at the early times of the collision.
As we explained in the Introduction, recently the phenomenology shifted to the ``fate of small initial state fluctuations",
related with the conical structure and ``ridges". Although we have not applied our results, it is clear 
 that such flows would be sensitive to higher gradients, as the size of those fluctuations is an order of magnitude smaller than
 the nuclear size  associated with  radial and elliptic flows studied before.
%We would like to emphasize  the potential role of the higher gradient terms on phenomenological
%applications. 
In agreement with our proposal \cite{LS}, the viscosity function (its real part) is a decreasing
function both of frequency and momenta. This behavior might be the reason behind the low viscosity observed at
RHIC. It may also explain the exceptionally good survival of various hydrodynamic flows, particularly the 
sound waves.

\section*{Acknowledgments}

We would like to express our gratitude for numerous discussion of the subject to  our Stony Brook colleagues Alexander Abanov,
Andrej Parnachev, Claudia Ratti, Shlomo Razamat, Derek Teaney, Izmail Zahed and 
Peter van Nieuwenhuizen. 

We  also wish to thank  
Chris Herzog, Alex Buchel, Maxim Khodas, Alex Kovner, Pavel Kovtun, Philip Mannheim, Alfred Mueller, Paul Romatschke,
Dam Son, Andrej Starinets, Misha Stephanov, Larry Yaffe, and Amos Yarom for very illuminating discussions
related to the present work.

We thank the [Department of Energy's] Institute for Nuclear Theory at the University of Washington for 
its hospitality and the Department of Energy for partial support during the completion of this work.

We thank the KITP at Santa Barbara and the GGI at Florence for the hospitality during the completion of this work. 
This research was supported in part by the National Science Foundation under Grant No. PHY05-51164.

\noindent This work is partially supported by the DOE grants DE-FG02-88ER40388 and DE-FG03-97ER4014.

\appendix

\section{Appendix: Analytic expansions of correlators} 
\label{sec:C}

In this Appendix we present  analytic expansions of retarded correlators at 
small frequency and momentum, as computed from the bulk gravity. 
The expansions are obtained following
 Appendix of Ref. \cite{BSSSR} where a perturbative approach to solving 
eq. (\ref{eq:master-equation}) is set.
We reproduce and extend their results to include some of   higher order terms. 

\subsection*{Scalar channel}

\begin{eqnarray}
A&=&1\,+\,i\,{\ln 2 \over 2}\,\omega\,+\,\ln2\,\left({3\,\ln 2\over 8}-\,1\right)\,\omega^2\,+\,\ln 2\,k^2\,-\,{\ln^2 2\over 2}\,k^4\,-\,
i\,{\ln^2 2\over 2}\,\omega\,k^2\,+\,\left({5\over 4}\,-\,{\ln^3 2\over 6}
\right)\,k^6\,\ldots\nonumber \\
B&=&{1\over 2}\,k^2\,+\,i\,{1\over 2}\,\omega\,+\,
\left({3\over 4}-{\ln 2\over 2}\right)\,k^4\,-\,\left({1\over 2}-{\ln 2\over 4}\right)\,
\omega^2\,-\,i\,{\ln 2\over 4}\,\omega\,k^2\,+\,
{\ln 2\over 4}\left(3-\ln 2\right)\,k^6\,\ldots 
\end{eqnarray}
For the retarded correlator $G^T$ we obtain
\beq
{1\over (\epsilon+P)}\,
G^T\,=\,-\,{B\over A}\,=\,-\,{1\over 2}\,k^2\,-\,i\,{1\over 2}\,\omega\,-\,{1\over 2}\,(\ln 2\,-\,1)\,
\omega^2\,-\,{1\over 4}\,\left(3\,-\,4\,\ln 2\right)\,k^4\,+\,i\,\ln 2\,\omega\,k^2\,-\,\ln^2 2\,k^6\,\ldots
\eeq

\subsection*{Shear channel}

\begin{eqnarray}
A&=&\omega\,+\,i\,{1\over 2}\,k^2\,+\,i\,{1\over 4}\, k^4\,+\,{\ln 2\over 4}\,w\,k^2\,+\,
i\,{\ln 2\over 2}\,\omega^2\ldots; \nonumber \\
B&=&{i\over 2}\, (k^2\,-\,\om^2)\,(1\,+\,i\,{2\,-\,\ln 2\over 2}\,\omega\,-\,{1\over 2}\,k^2\ldots)
\end{eqnarray}
The correlator reads
\beq
{1\over (\epsilon+P)}\,
G^D\,=\,-\,{k^2\over \om^2\,-\,k^2}\ {B\over A}\,=\,{i\,k^2/2\,[1\,+\,i\,(2\,-\,\ln 2)\,\om\,-\,k^2/2\,\ldots]\,+\,\om\,k^2/2\,+\cdots
\over\,\om\,+\,i\,k^2/2\,[1\,+\,i\,(2\,-\,\ln 2)\,\om\,-\,
k^2/2+\,\cdots]}
\eeq
\subsection*{Sound channel}

\begin{eqnarray}
&&A\,=\,8\,\left[9\,\omega^2\,-\,3\,k^2\,+\,i\,6\,\omega\,k^2\,+\,(\ln 2\,-\,4)\,k^4\,+\,
9\,\ln2\,({\ln 2\over 2}\,-\,1)\,\omega^4\,+\,3\,\ln 2\,(2\,-\,{\ln 2\over 2})\,\omega^2\,k^2
\right]\nonumber \\
&&B\,=\,3\,\left[ 18\,\omega^2\,-\,30\,k^2\,+\,i\,12\,\omega\,k^2\,+\,2\,(5\,\ln 2\,-\,12)\,k^4\,+\,
3\,\ln 2\,(12\,-\,5\,\ln 2)\,\omega^2\,k^2\,+\,\right.\nonumber \\
&&\ \ \ \ \ \ \ \ \ \ \ \ \ \ \ \ \ \ \ \ \ \ \ \ \ \ \ \ \ \ \ \ \ \ \ \ \ \ \ \ \ \ \ \ \ \ \ \ \ \ \ \ \ \ \ \ \ \ \ \ \ \ \ \ \ \ \ \ \ \ \ \ \ \ \ \ \ \ \ \ \ \ \ \ \ \ 
+\,\left. 9\,\ln 2\,(\ln 2\,-\,2)\,\omega^4\,\ldots
\right]
\end{eqnarray}
The analytically controlled part of the sound correlator 
\begin{eqnarray}
{1\over (\epsilon+P)}\,G^S&=& \left({4\over 3}{k^4\over (\om^2\,-\,k^2)^2}\,{B\over A}\,+\,{1\over 12}\,{29\,k^4\,-\,30\,k^2\,\om^2\,+\,
9\,\om^4\over (k^2\,-\,\om^2)^2}\,-\,{3\over 4}\right)\,{\omega^2\over k^2}\,+\,1
\nonumber \\
&=&\frac{-\,k^2 \,+\,i\,2\,[1\,-\,i\,\omega\,(\ln 2\,-\,2)\,+\,\cdots]\,\omega\,k^2\,+\,2\,\omega^2\,k^2\,\ldots}
{3\,\omega^2\,-\,k^2\,+\,i\,2\,\omega\,k^2\,[1\,-\,i\,\omega\,(\ln 2\,-\,2)\,+\,\cdots]}
\end{eqnarray}
Here we used eq. (\ref{shift}).

\section{Appendix: Correlators from generalized
hydrodynamics} \label{sec:B}

In this Appendix we compute the retarded correlators from the hydrodynamic ansatz (\ref{Pi})
using 4d metric perturbations. The non-perturbed space
has the Minkowski metric $g^{\mu\nu}\,=\,diag\{-1,1,1,1\}$.

\subsection*{Scalar channel}

The perturbation is $h\equiv h_{xy}(z,t)$. The fluid remains at rest.
We first compute Christoffels coefficients and Riemann tensor
\begin{eqnarray}
&&\Gamma^t_{xy}\,=\,\Gamma_{ty}^x\,=\,\Gamma^y_{tx}\,=\,{1\over 3}\,\dot h\,; \ \ \ \ \ \ \ \ \ \ \ 
\Gamma^x_{zy}\,=\,\Gamma_{zx}^y\,=\,-\,\Gamma_{xy}^z\,=\,{1\over 2}\,h^\prime\nonumber \\
&& R^t_{xty}\,=\,R^x_{tty}\,=\, R^y_{ttx} \, =\,{1\over 2}\,\ddot h\,; \ \ \ \ \ \ \ \ \ \ \ 
R^x_{zzy}\,=\,R^y_{zzx}\,=\,-\,R^z_{xzy}\,=\,{1\over 2}\,h^{\prime\prime} \,;\nonumber \\
&&R^t_{xzy}\,=\,R^x_{tzy}\,=\,R^y_{tzx}\,=\,-\,R^z_{xzy}\,=\,R^x_{zty}\,=\,R^y_{ztx}\,=\,{1\over 2}\,\dot h^\prime
\end{eqnarray}
The only non-zero component of the Ricci tensor is $R_{xy}$ while the scalar curvature is zero
\beq
R_{xy}\,=\,{1\over 2}\,(\ddot h \,-\,h^{\prime\prime})\ \ \ \ \ \ \ \ \ \ \ \ \ \ \ \ R\,=\,0
\eeq
The relevant non-zero components of the Weyl tensor are
\begin{eqnarray}
&&C_{txty}\,=\, -\,C_{txyt}\,=\,C_{xtyt}\,=\,C_{ytxt}\,=\,C_{xzyz}\,=\,C_{yzxz}=\,-\,{1\over 4}\,(\ddot h \,+\,h^{\prime\prime}) \nonumber \\
&&C_{xzyt}=\,C_{yzxt}\,=\,C_{xtyz}\,=\,C_{ytxz}\,=\,-\,{1\over 2}\,\dot h^\prime
\end{eqnarray}
The $xy$ component of the stress tensor reads
\beq
\langle T^{xy}\rangle\,=\,-\,P\,h\,-\,\eta\,\dot h \,-\, {1\over 2}\,\kappa\,[\ddot h \,+\,h^{\prime\prime}]\,+\,
\rho\,{1\over 2}\,[ \stackrel{{\mathbf .} \;\!\!. \;\!\! .}{h} \,- \,\dot h^{\prime\prime}]\,-\,
\xi\,{1\over 4}\,[\stackrel{{\mathbf .} \;\!\!. \;\!\! .\;\!\!.}{h} \,-\,2\,\ddot h^{\prime\prime}\,+ \,h^{\prime\prime\prime\prime}]
\eeq
In momentum space this becomes
\beq\label{Txy}
\langle T^{xy}\rangle \,=\,-\,\left[P\,-\,i\,\omega\,\eta\,-\,\kappa\, {1\over 2}\,(w^2 \,+\,k^2)\,-\,
\rho\,{i\,\omega\over 2}\,(w^2 \,- \,k^2)\,+\,
\xi\,{1\over 4}\,(\omega^2\,-\,k^2)^2\right]\ h(k,w)
\eeq
From Eq. (\ref{Txy}) one can read off the correlator $G^T\,=\,G^{xyxy}$
\beq
\tilde G^{xyxy}(k,w)\,=\,P\,-\,i\,\omega\,\eta\,-\, \kappa\,{1\over 2}\,(w^2 \,+\,k^2)\,-\,
\rho\,{i\,\omega\over 2}\,(w^2 \,- \,k^2)\,+\,
\xi\,{1\over 4}\,(\omega^2\,-\,k^2)^2
\eeq
The retarded correlator $G^{xyxy}\,=\,\tilde G^{xyxy}\,-\,P$.
The transverse static susceptibility $\chi^T$ is momenta dependent and is given by 
the functions $\kappa$ and $\xi$:
\beq
\chi^T(k)\,=\,\,-\,\kappa(k,0)\,k^2/2\,+\,
\xi(k,0)\,\,k^4/4
\eeq

\subsection*{Shear channel}

The perturbation $h\equiv h_{tx}(z,t)$. The fluid's four velocities is
$
u^\mu\,=\,(1,v,0,0)$ and $u_\mu\,=\,(-1, v+h,0,0)$
The Christoffels coefficients and the Riemann tensor are 
\begin{eqnarray}
&&\Gamma^{x}_{tz}\,=\,-\,\Gamma^{t}_{xz}\,=\,-\,\Gamma^{z}_{xt}\,=\,{1\over 2}\,h^\prime\,; \ \ \ \ \ \ \ \ \ \ \ \ \ 
 \Gamma^x_{tt}\,=\,\dot h\nonumber \\
&&R^t_{txtz}\,=\,-\,R^t_{xzt}=\,R^x_{ttz}\,=\,-\,R^x_{tzt}\,=\,-\,R^z_{txt}\,=\,-{1\over 2}\,\dot h^\prime\nonumber \\ 
&&R^z_{xzt}\,=\,-\,R^z_{xtz}\,=\,R^x_{ztz}\,=\,-\,R^z_{txz}\,=\,-\,{1\over 2}\, h^{\prime\prime}
\end{eqnarray}
The non-zero Ricci components and curvature are
\beq
R_{xz}\,=\,-{1\over 2}\,\dot h^\prime\,;\ \ \ \ \ \ \ \ \ \ \ \ 
R_{xz}\,=\,-{1\over 2}\,h^{\prime\prime}\,;\ \ \ \ \ \ \ \ \ \ \ \ R\,=\,0
\eeq
The relevant non-zero Weyl components read
\beq
C_{xtzt}\,=\,C_{ztxt}\,=\,{1\over 4}\,\dot h^\prime\,;\ \ \ \ \ \ \ \ \ \ \ \ 
C_{xzzt}\,=\,C_{ztxz}\,=\,{1\over 4}\, h^{\prime\prime}
\eeq
The components of the stress tensor 
\begin{eqnarray}
&&\langle T^{tt}\rangle \,=\,\epsilon\,;\ \ \ \ \ \ \ \ \ T^{tx}\,=\,(\epsilon\,+\,P)\,v\,+\,P\,h\,\nonumber \\
&&\langle T^{xz}\rangle
\,=\,-\,\eta\,v^\prime\,+\,\kappa\,{1\over 2}\,\dot h^\prime\,+\,\rho\,{1\over 4}\,(h^{\prime\prime\prime}\,-\,2\,
\ddot h^\prime)\,+\,\xi\,{1\over 4}\,(\stackrel{{\mathbf .} \;\!\!. \;\!\! .}{h}^\prime\,-\,\dot h^{\prime\prime\prime})
\end{eqnarray}
Equations of motion relate the metric perturbation $h$ to the induced three-velocity $v$:
\begin{eqnarray}
&&\partial_t\,\langle T^{tx}\rangle\,=\, (\epsilon\,+\,P)\,(\dot v\,+\,\dot h)\,;\nonumber \\  
&&\partial_z\,\langle T^{zx}\rangle\,=\,-\,\eta\,v^{\prime\prime}\,+\,\kappa\,{1\over 2}\,\dot h^{\prime\prime}
\,+\,\rho\,{1\over 4}\,(h^{\prime\prime\prime\prime}\,-\,2\,
\ddot h^{\prime\prime}\,+\,\xi\,{1\over 4}\,(\stackrel{{\mathbf .} \;\!\!. \;\!\! .}{h}^{\prime\prime}
\,-\,\dot h^{\prime\prime\prime\prime})
\end{eqnarray}
which leads to the relation (in momentum space)
\beq
v\,=\,h\,\frac{i\,\omega\,-\,i\bar\kappa\,\omega\,k^2/2\,-\,\bar\rho\,k^2\,(k^2\,-\,2\,\omega^2)/4\,+
\,i\,\bar\xi\,\omega\,k^2\,(\omega^2\,-\,k^2)/4}{-i\,\omega\,+\,\bar\eta\,k^2}
\eeq
Substituting this relation back into the expression for $\langle T^{tx}\rangle$ we can read off the correlator $G^D\,=\,G^{txtx}$
\beq
\tilde G^{txtx}\,=\,(\epsilon\,+\,P)\ 
\frac{\bar\eta\,k^2\,-\,i\bar\kappa\,\omega\,k^2/2\,-\,\bar\rho\,k^2\,(k^2\,-\,2\,\omega^2)/4\,+
\,i\,\bar\xi\,\omega\,k^2\,(\omega^2\,-\,k^2)/4}{-i\,\omega\,+\,\bar\eta\,k^2}\,-\,\epsilon
\eeq
Note the appearance of the extra terms proportional to the GSFs in the numerator, 
whereas in the normal diffusion scenario the residue is usually given by the viscosity only.

The retarded correlator $G^{txtx}\,=\,\tilde G^{txtx}\,+\,\epsilon$.
The shear static susceptibility $\chi^D$:
\beq
\chi^D(k)\,=\,(\epsilon\,+\,P)\,\left[1 \,-\,{\bar \rho(k,0)\over 4\,\bar \eta(k,0)}\,k^2\right]
\eeq

\subsection*{Sound channel}

The perturbation which generates sound is $h\equiv h_{tz}(z,t)$. 
The fluid`s four velocities is $u^\mu\,=\,(1,0,0,v)$ and $u_\mu\,=\,(-1,0,0, v+h)$.

Christoffels and Riemann are
\beq
\Gamma^{t}_{zz}\,=-\,\,h^\prime\,; \ \ \ \ \ \ \ \ \ \ \ \ \ 
 \Gamma^z_{tt}\,=\,\dot h\,; \ \ \ \ \ \ \ \ \ \ \ \ \ 
R^z_{tztz}\,=\,-\,R^z_{ttz}=\,R^t_{zzt}\,=\,-\,R^t_{ztz}
\eeq
Contrary to the cases of tensor and shear perturbations, the sound perturbation has 
a nonvanishing scalar curvature.
\beq
R_{zz}\,=\,-\,R_{tt}\,=\,-\,\dot h^\prime\,;\ \ \ \ \ \ \ \ \ \ \ \ \ \ \ \ \ \ 
R\,=\,-\,2\,\dot h^\prime\,;\ \ \ \ \ \ \ \ \ \ \ \ \ \ \ \ \ \
C_{ztzt}\,=\,{1\over 3}\,\dot h^\prime
\eeq
The relevant components of the stress tensor
\begin{eqnarray}
&&\langle T^{tt}\rangle\,=\,\epsilon\,; \ \ \ \ \ \ \ \ \  \ \ 
\langle T^{tz}\rangle\,=\,(\epsilon\,+\,P)\,v\,+\,P\,h\,;\nonumber \\
&&\langle T^{zz}\rangle
\,=\,P\,-\,\eta\,{4\over 3}\,v^\prime\,+\,\kappa\,{2\over 3}\,\dot h^\prime\,-\,\rho\,{2\over 3}\,\ddot h^\prime
\,+\,\xi\,{1\over 3}\,\stackrel{{\mathbf .} \;\!\!. \;\!\! .}{h}^{\prime}
\end{eqnarray}
%where 
%\beq
%\eta\,=\,\eta_1\,+\,\eta_2\,(\omega^2\,-\,2\,k^2)
%\eeq
Equations of motion can be solved for $v$ relating it to the perturbation $h$
\beq
v\,=\,h\,\frac{3\,\omega^2\,-\,2\,\bar\kappa\,\omega^2\,k^2\,-\,2\,i\,\bar\rho\,\omega^3\,k^2\,+\,\bar\xi\,\omega^4\,k^2}
{k^2\,-\,3\,\omega^2\,-\,4\,i\,\bar\eta\,\omega\,k^2}
\eeq
Substituting $v$ back into the expression for $T^{tz}$ we can read off the the correlator $G^S\,=\,G^{tztz}$
\beq
\tilde
G^{tztz}\,=\,(\epsilon\,+\,P)\ \frac{k^2\,-\,4\,i\,\bar\eta\,\omega\,k^2\,
-\,2\,\bar\kappa\,\omega^2\,k^2\,-\,2\,i\,\bar\rho\,\omega^3\,k^2\,+\,\bar\xi\,\omega^4\,k^2}
{k^2\,-\,3\,\omega^2\,-\,4\,i\,\bar\eta\,\omega\,k^2}\,-\,\epsilon
\eeq
The retarded correlator $G^{tztz}\,=\,\tilde G^{tztz}\,+\,\epsilon$

\end{document}